\documentclass[epj]{svjour}

\usepackage{graphics}
\usepackage{mathrsfs}

\begin{document}

\title{Comparison of pure and combined search strategies for single and multiple
targets}
\subtitle{}
\author{Vladimir V. Palyulin\inst{1} \and Vladimir N. Mantsevich\inst{2}
\and Rainer Klages\inst{3} \and Ralf Metzler\inst{4} \and Aleksei V.
Chechkin\inst{4,5}}
\institute{Department of Chemical Engineering and Biotechnology, University of
Cambridge, Cambridge CB3 0AS, United Kingdom
\and Physics Department, Moscow State University, Moscow, 119991, Russia
\and Queen Mary University of London, School of Mathematical Sciences,
Mile End Road, London E1 4NS, United Kingdom
\and Institute for Physics \& Astronomy, University of Potsdam, D-14476
Potsdam-Golm
\and Akhiezer Institute for Theoretical Physics NSC KIPT, Kharkov, 61108, Ukraine}

\date{Received: date / Revised version: date}

\abstract{We address the generic problem of random search for a point-like
target on a line. Using the measures of search reliability and efficiency
to quantify the random search quality, we compare Brownian search with
L{\'e}vy search based on long-tailed jump length distributions. We then
compare these results with a search process combined of two different long-tailed
jump length distributions. Moreover, we study the case of multiple targets
located by a L{\'e}vy searcher.
}

\maketitle

\section{Introduction}
\label{intro}

Searching for randomly located targets is a central problem in many branches of
the sciences comprising all scales from the smallest to the largest: Examples
include chemical reactions, in which a molecule has to find a reactive target
such as the search of transcription factor proteins for a specific binding spot
on a DNA chain \cite{mirny2007,max2013}, the question of molecular signal
detection \cite{berg,bialek,levine,aljaz}, white blood cells trying to locate
intruding pathogens \cite{HaBa12}, spider monkeys searching for food in a tropical
forest \cite{RFM03}, human rescue operations \cite{rescue}, the hunt for submarines
\cite{Shles06} and, more mathematically, algorithms for finding the minima in a
complex search space \cite{pavlyukevich}.

In society the development of search strategies like the search for land mines,
castaways or victims of avalanches belongs to the realm of {\em operations
research\/} \cite{Stone07,AG03}. In ecology and biology understanding the
foraging of biological organisms forms part of the new discipline of {\em
movement ecology\/} \cite{Nath08,MCB14}. Prominent examples for the latter are
wandering albatrosses searching for food \cite{Vis96,Vis99,Edw07}, marine
predators diving for prey \cite{Sims08,Sims10}, and bees collecting nectar
\cite{LICCK12}. Within this context the {\em L\'evy Flight Hypothesis} (LFH)
attracted considerable attention \cite{shlesinger86}: It predicts that under
certain mathematical conditions scale-free jump processes called {\em L\'evy
flights} (LFs) \cite{SZK93} minimise the search time \cite{Vis96,Vis99,VLRS11}.
The LFH implies that, for instance, for a bumblebee searching for rare flowers
the flight lengths should be distributed according to a power law. This prediction
is completely different from the paradigm put forward by Pearson more than a
century ago who proposed to model the movements of biological organisms by simple
random walks \cite{Pea06}. Pearson's theory entails that the movement lengths are
distributed according to a Gaussian distribution, contrasting the L{\'e}vy stable
form underlying the LFH. Both L\'evy and Gaussian dynamics represent fundamentally
different, {\em pure\/} classes of stochastic processes.

However, in complex biological reality animals, or humans, may not search
according to a simple, pure stochastic process, as other factors may come into
play. For example, they often have a limited perception while moving with high
speed. In this case a more promising search strategy is to switch between a slow
recognition mode during which targets can be found and fast relocations
\cite{AmSci1990,AmerZool2001,BenichouIntermittent}. These {\em intermittent search
strategies\/} can {\em combine\/} different types of motion such as Brownian,
ballistic motion, or LFs \cite{benichou2005,benichou2008,ReynoldsPhysA2009,gleb1,%
gleb2,prl2005}. They may also include various distributions for the switching times
from one phase to another \cite{LomholtPNAS2008}. The optimal search strategy
then depends on the specific types of motion and the dimension of the search space
\cite{BenichouIntermittent,LomholtPNAS2008,benichou2006,benichou2007}.

For real world problems it is furthermore crucial that a searcher does not only
eventually find a target but also that the search is successful within a limited
time span, for instance, if the search is a rescue operation or if a starving
animal searches for food to survive. This means that the search needs to be {\em
efficient}. However, on top of this it often must also be {\em reliable\/} in
that a given target is not missed out but found with sufficiently high
probability \cite{PNAS14,LevyLong}.

\begin{figure}
\resizebox{0.49\textwidth}{!}{\includegraphics{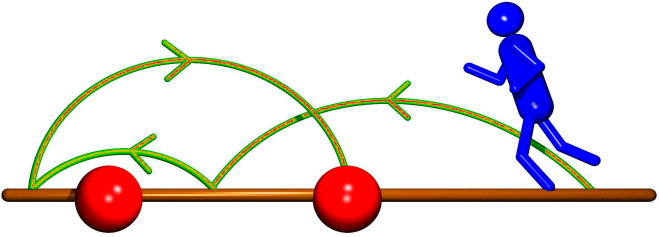}}
\caption{L\'evy searcher multiply hopping across (overshooting) the two targets
before eventually hitting the right target.}
\label{schematic}
\end{figure}

From a mathematical point of view, finding a single target when hitting it (in
the attempt of sweeping past it) defines a {\em first passage problem\/}
\cite{redner,MetzlerOshaninRedner}, which corresponds to the situation where a
searcher searches uninterruptedly while continuously moving. On the other hand,
landing on a single target after a relocation process has been completed can be
formulated as a {\em first arrival problem\/} \cite{PNAS14,LevyLong,JPA2003}.
This distinction relates to the situation described above, when a searcher does
not perceive targets while moving but only if it comes at rest exactly on the
target, or sufficiently close to it (see Fig.~\ref{schematic}). Calculating
first passage and first arrival
times for stochastic processes are well-defined mathematical problems, that, in
certain cases such as for Brownian motion and in one dimension, can be solved
exactly. Remarkable recent progress has been reported for first passage problems,
such as the universality of certain classes of mean and global mean first passage
times \cite{benichou2008,yossinat,chevalier}. Moreover, it has been discussed that
mean search times are not
always meaningful, as they may not be representative \cite{gleb,thiago}, or that
they are vastly different from the most probably first passage time obtained from
the full distribution of first passage times \cite{aljaz1,aljaz2}.
However, these studies only address the problem of finding {\em single\/} targets.
For solving the problem of how to find {\em multiple\/} targets other techniques
are needed. The extreme case of the time needed to find all targets in a given
domain with certainty is called the {\em covering time\/} \cite{aldous}. Recent
work has demonstrated that for a broad range of stochastic processes on networks
there holds a certain universality for the distributions of cover times
\cite{CBV15}. Clearly, first passage as well as arrival problems for finding
single targets and cover times for finding all of them define extreme cases of
search problems. We here consider the case of a finite number of targets and
explicitly calculate the splitting probabilities to locate one of the targets.
This setting is distinct from the previously studied case of equally spaced
targets in a setting with periodic boundary conditions \cite{LomholtPNAS2008,epl}.

The structure of our paper is as follows: In Section 2 we briefly review two
important applications of search theory to biology. The first one is the problem
to understand the foraging of biological organisms. Within this context we also
elaborate on the role of {\em L\'evy walks\/}, which define a special case of
{\em continuous time random walks}, that in turn are the central topic of this
Special Issue. The second one is the problem of search along DNA chains by DNA
binding proteins. Section 3 starts by introducing two basic quantities for
judging the quality of a mathematical search problem, namely, the search
reliability and the search efficiency. In Section 4 these two quantities are
calculated explicitly for the single-target first arrival problem of the two
fundamental stochastic processes of pure Brownian motion and a pure LF search
process, the latter both with and without a bias. Here we review known results in
the literature. Section 5 generalises this theory to single-target search by a
combination of two L\'evy stable processes, which yields a new result. In Section 6
we address the problem of search for more than one target by a pure stochastic
process.  As a specific example, we consider Brownian and L\'evy search of two
targets by calculating the first arrival density, the search reliability, and the
search efficiency. We conclude with a brief outlook in Section 7. In the Appendix we
collect a number of technical results.

\section{Search research: two examples from biology}

\subsection{Search for food by biological organisms}

The advantage of random search based on random walks with long-tailed, scale-free
jump length distributions was postulated by Shlesinger and Klafter already in 1986
\cite{shlesinger86}. The groundbreaking moment for the popularisation of this
concept came with the 1996 article by Viswanathan and colleagues: In this study
the flight times of albatrosses were recorded during their foraging excursions
in the South Atlantic \cite{Vis96}. It was found that the distribution of flight
times obeyed an asymptotic power law $\sim t^{-2}$. Assuming that the birds move
with a constant average speed one can associate these flight times with the
respective power law distribution $\approx|x|^{-2}$ of flight lengths. This
suggests that the albatrosses were searching for food by performing \emph{L\'evy
flights\/}. For more than a decade albatrosses were thus considered to be the
most prominent case study of animal foraging by LFs. This work
spawned a large number of related studies suggesting that many other animals
like goats and deer \cite{goat,deer}, bumblebees \cite{Vis99}, spider monkeys
\cite{RFM03}, marine predators \cite{Sims08,Sims10}, and micro-zooplankton
\cite{dinoflagellate} also perform L\'evy search \cite{VLRS11}. Heavy-tailed
distributions were also found to be characteristic for human movement dynamics
\cite{brockmann1,brockmann2}. Interestingly, the discussion of the LF nature of the
flight of the albatross recently saw an interesting twist. While a re-analysis of
the albatross flights showed that they generally are \emph{not\/} LFs
\cite{albatros}, strong evidence was presented according to which LFs are indeed
a search pattern for individual albatrosses \cite{PNAS2012}.

The mathematical underpinning for these relevance of long-tailed probability
laws was provided in the works starting with L{\'e}vy \cite{Levy37} as well as
Gnedenko and Kolmogorov \cite{GnKo54}. Their work showed that specific types of
power laws, the {\em L\'evy alpha stable distributions} \cite{Levy37,bouchaud,%
hughes,KRS08}, obey a generalised central limit theorem. Their result thus
generalises
the conventional central limit theorem for Gaussian distributions, which explains
why Brownian motion with a Gaussian probability distribution is universally
observed in a huge variety of physical phenomena. But, Gaussian tails decay
faster than power laws, which implies that for L\'evy-distributed flight lengths
there is a larger probability to yield long flights than for flight lengths
obeying Gaussian statistics. The generalised central limit theorem then
guarantees that for sufficiently many steps in the L{\'e}vy flight a well
defined limit distribution for the probability distribution emerges. The only
difference is that this L{\'e}vy stable law is not universal as the Gaussian, but
characterised by a specific alpha stable index \cite{Levy37,bouchaud,hughes,KRS08}.
Intuitively, L\'evy flights should be better suited to scan a large space for
randomly distributed targets than Brownian motion. In turn Brownian search should
outperform L\'evy motion when the targets are dense. This is the basic idea
underlying the LFH mentioned above \cite{Vis99}. The motivating questions are:
{\em What is the best statistical strategy to adapt in order to search efficiently
for randomly located, sparse objects?} The LFH stipulates that {\em L\'evy motion
provides an optimal search strategy for sparse, randomly distributed, immobile,
revisitable targets in unbounded domains.} \cite{Vis99,VLRS11}.

To be precise we note that there exists two formulations of continuous time
random walk processes with long-tailed, scale-free relocation distributions.
One pertains to L{\'e}vy flights, these are fully Markovian processes in which
the jumps occur instantaneously, separated with a well defined sojourn time.
In exchange the mean squared displacement of L{\'e}vy flights diverges
\cite{bouchaud,hughes,KRS08,MeKl00}. This divergence is remedied in the \emph{
L{\'e}vy walk\/} model, in which a spatiotemporal coupling between relocation
lengths and waiting times exists, such that long jumps are penalised by long
waiting times \cite{KRS08,MeKl00,wong,ZaburdaevReview,KlSo11}. In fact a
specific L{\'e}vy walk model was investigated by computer simulations in the
analysis leading to the LFH \cite{Vis99}. Remarkably, in the limit of sparse
food both L{\'e}vy walk and flight dynamics lead to the same optimal L{\'e}vy
stable exponent $\alpha=1$ for the distribution of relocation lengths
\cite{Vis99,prl2005}. We also note that a rigorous mathematical proof of the
LFH to date remains elusive, while empirical tests are debated in literature
\cite{Edw07,Buch08,dJWH11,JPE11,Pyke15,Reyn15}.

\subsection{Search along DNA chains}

To activate or downregulate individual genes on the genome, in biological cells
specific DNA-binding proteins needs to locate and then bind to designated binding
sites on the DNA chain. For long it had been assumed that a good estimate for
the associated binding rate is the celebrated Smoluchowski result for molecular
aggregation \cite{smoluchowski}. In vitro experiments for the search rate of
the Lac repressor protein remarkably showed a rate, that was larger by around
two orders of magnitude \cite{lac}. Building on earlier work of Adam and
Delbr{\"u}ck \cite{adam} and Richter and Eigen \cite{richter}, the so-called
facilitated diffusion model was developed by Berg and von Hippel, and coworkers
\cite{Berg1977,Berg1981,Berg1989}. The main idea of the facilitated diffusion
model is the possibility that the searching protein may not only diffuse in
the bulk volume of the reaction container, but it may also intermittently
associate with the DNA chain and perform a random sliding motion on it. Hereby
the linear topology of the DNA leads to a transient dimensional reduction of
the random search, effecting a similar advantage as the intermittent search
model discussed above. Namely, the one-dimensional search makes sure that the
target, if close-by, will be located with high probability. Significant
oversampling due to the recurrent motion of one-dimensional diffusion is
avoided by the intermittent volume excursions, that decorrelate the position
of the protein before it rebinds to the DNA. Indeed, this approach to good
approximation explains the observed speedup compared to the Smoluchowski limit,
see the recent review \cite{BenichouPhysRep2012}.

In single molecule measurements the various search modes can be verified directly
or indirectly, for instance, the existence of the one-dimensional sliding motion
\cite{SlidingExp,WangAustinCoxPRL2006,mark}, association-dissociation events
leading to the change between sliding and bulk diffusion \cite{AssDissExp},
intersegmental transfer between different segments of DNA \cite{IntersegExp},
and the role of the three-dimensional DNA conformations \cite{gijs1}. A
number of theoretical studies highlight the role of the intermittency of the
search for the efficiency of the process. Thus, Halford and Marko
\cite{HalfordMarko2004}, Coppey et al. \cite{Benichou2004}, Erskine et al.
\cite{Erskine}, Givaty and Levy \cite{LevyJMB2009}, and Klenin et al.
\cite{Klenin} considered the competition of one- and three-dimensional
diffusion. Slutsky and Mirny \cite{SlutskyMirny} argue that the one-dimens\-ional
search needs to consist of search and recognition modes such that the protein
can slide sufficiently fast while retaining its binding selectivity. Including
intersegmental transfers or jumps of the protein between chemically remote but
physically close segments of the DNA chain further improves the search efficiency,
especially when the three-dimensional search mode is repressed, for instance, at
certain salt conditions \cite{prl2005,gijs2,shklovskii2007}. Interestingly, in
the limit of long DNA chains and sufficiently fast reorganisation of the DNA
conformation, intersegmental jumps effect an LF search by the binding protein
\cite{prl2005}. In fact, this may be the only example for an LF, which is not
hampered by a diverging second moment: as the jumps are long-tailed in terms
of the chemical distance measured along the backbone of the DNA molecule, but
are local in the real, embedding space, this divergence is physically meaningful.

Intersegmental jumps
may even assist in avoiding "roadblocks" in the form of other non-specifically
DNA-bound proteins \cite{LiBerg2009,markovitz2013}. A concise overview over the
various facilitated diffusion search modes is given in \cite{max2012}. More recent
progress includes the formulation of facilitated diffusion for the in vivo case
of living bacteria cells \cite{max2013} and the inclusion of effects of the DNA
sequence \cite{tolyaspeedselectivity2013,larson2013,max_sequence}. Finally, effects
of the crowded cytoplasm of living cells were considered by different approaches
in \cite{spakowitz,liu}. A path integral formulation of the downregulation of one
gene by the product of a steering gene, including the stochasticity of the
regulation process \cite{noise} was given in \cite{otto}. A noteworthy result of
that study is that the efficiency of the protein search for its target binding
site crucially depends on the initial distance from this target \cite{otto}, a
result that is consistent with the so-called rapid search hypothesis based on
bioinformatics research \cite{mirny2007}.

We note that while there exists full experimental evidence for the intermittent
search of DNA binding proteins, the showcase examples of Lac repressor proteins
or EcoRV restriction enzymes are in fact rather untypical, and many proteins
simply occur at sufficiently high concentrations and utilise pure three-dimensional
diffusion to locate their binding site \cite{HalfordReview2009,finkelstein2012}.
Yet for those proteins whose number per cell is small, facilitated diffusion is
essential \cite{berg2012}.

\section{Defining search reliability and search efficiency}

As mentioned above, the key quantities to characterise the success of a search
strategy are the reliability and efficiency. The former quantifies the probability
whether the search process is ever successful, the latter is a measure for how
long the search takes. We define the search reliability as the cumulative
probability $P$ of the first arrival to reach the target. In terms of the survival
probability $\mathscr{S}(t)$ (of not hitting the target up to time $t$), we thus
write $P=1-\mathscr{S}(\infty)$ \cite{redner,MetzlerOshaninRedner}. Expressing the
survival probability in terms of the first arrival time density $\wp_{\mathrm{fa}}
(t)$, we thus
find the relation
\begin{equation}
P=1-\mathscr{S}(\infty)=\int_0^{\infty}\wp_{\mathrm{fa}}(t)dt.
\end{equation}
Using the Laplace transform, defined through
\begin{equation}
\wp_{\mathrm{fa}}(s)=\int_0^\infty\wp_{\mathrm{fa}}(t)e^{-st}dt,
\end{equation}
we find the relation \cite{PNAS14}
\begin{equation}
P=\lim_{s\rightarrow0}\wp_{\mathrm{fa}}(s).
\label{Pdef}
\end{equation}
The search reliability depends on the exact type of the random search process
as well as the geometrical details (dimension, distance from the starting
position to the target etc.). The arrival time density $\wp_{\mathrm{fa}}(t)$
can be determined from the (fractional) Fokker-Planck equation of the search process,
equipped with a sink term \cite{PNAS14,LevyLong,JPA2003}.

For search in one dimension by LFs without a bias the search reliability is
unity if $\alpha>1$ and zero otherwise \cite{PNAS14,LevyLong,JPA2003}, which is
consistent with previous results \cite{spitzer}. For search in the presence of
a bias (water stream for marine searchers, winds for airborne foragers, etc.)
the search reliability can vary between zero and unity \cite{PNAS14,LevyLong},
which is true also for Brownian motion \cite{redner}: when the bias pushes a
searcher away from the target the search reliability is exponentially suppressed
by a Boltzmann-like factor \cite{redner}. A search reliability of unity does not
necessarily imply recurrence of the motion. For instance, LFs with $\alpha=1$ in
one dimension and Brownian motion in two dimensions are recurrent but their search
reliability is zero.

The second quantity of interest is the search efficiency. Most of the
theoretical studies consider a probabilistic searcher with a limited radius of
perception. Motivated by \cite{james2010}, in this case two basic definitions
of the search efficiency are considered to be either
\begin{equation}
\mathrm{Efficiency}_1=\frac{\mathrm{visited\,number\,of\,targets}}
{\mathrm{number\,of\,steps}},
\label{eff1}
\end{equation}
or
\begin{equation}
\mathrm{Efficiency}_2=\frac{\mathrm{visited\,number\,of\,targets}}{
\mathrm{distance\,travelled}}.
\label{eff2}
\end{equation}
The first definition applies especially to {\em saltatory search}, where a
searcher moves in a jump-like fashion and is able to detect the target only
around the landing point after a jump. The second definition is adapted to
{\em cruise motion}, during which the searcher keeps exploring the search space
continuously during the whole search process. An example for the former scenario
is given by a regulatory protein that moves in three-dimensional space and
occasionally binds to the DNA of a biological cell until it finds its binding
site. The latter scenario would correspond to an eagle or vulture whose excellent
eyesight permits them to scan their environment for food during their entire
flight. For LFs, Eq.~(\ref{eff1}) presents a natural choice while Eq.~(\ref{eff2})
is better suited for processes like Brownian motion and finite-velocity L\'evy
walks.

In what follows we focus on the limit of sparse targets. Concretely, we consider
a single or a finite number of targets. For a single target and saltatory motion
we argued that the efficiency should be defined from Eq.~(\ref{eff1}) with proper
averaging \cite{PNAS14,LevyLong}. In our continuous time model the number of steps
is naturally substituted by the time of the process. We choose the following
averaging \cite{PNAS14,LevyLong}
\begin{equation}
\mathcal{E}=\left<\frac{1}{t}\right>=\int^\infty_0 \wp_\mathrm{fa}(s)ds
\label{Edef}
\end{equation}
over the inverse search times. This choice appears more meaningful than taking
an average of the form $1/\langle t\rangle$, as the latter would produce a zero
efficiency when the mean search time diverges. Our definition (\ref{Edef}) instead
pronounces short and intermediate search times.

\section{Search of a single target by a single L\'evy flight searcher}

Below we use the search reliability and efficiency to characterise search
strategies of the motion governed by two L\'evy stable processes and search by a
single L\'evy stable process for more than one target. Before, we recall the main
properties of the search of a single L{\'e}vy searcher in an environment
without and with an external bias, as well as the limit of a Brownian searcher.

The properties of an LF search process can be calculated from a space-fractional
Fokker-Planck diffusion equation \cite{MeKl00} for the non-normalised density
function $f(x,t)$ \cite{JPA2003},
\begin{equation}
\frac{\partial f(x,t)}{\partial t}=K_\alpha\frac{\partial^{\alpha}f(x,t)}{\partial
\left\vert x \right\vert^{\alpha}}-\wp_{\mathrm{fa}}(t)\delta(x-x_1),
\label{SinkFFPE}
\end{equation}
where the target, represented as a $\delta$-sink, is located at $x=x_1$. The
generalised diffusion coefficient has physical dimensions $[K={\alpha}]=\mathrm{
cm}^{\alpha}/\mathrm{sec}$. We
assume that at $t=0$ the searcher is placed at $x=x_0$, that is, $f(x,0)=
\delta(x-x_0)$. The $\delta$-sink effects the condition $f(x_1,t)=0$
\cite{JPA2003,prl2005}. In Eq.~(\ref{SinkFFPE}) the fractional derivative
$\partial^\alpha/\partial|x|^\alpha$ is conveniently represented in terms of its
Fourier transform \cite{MeKl00}
\begin{equation}
\int^{\infty}_{-\infty}e^{ikx}\frac{\partial^\alpha}{\partial|x|^\alpha}
f(x,t)dx=-|k|^\alpha f(k,t).
\end{equation} 
Integrating over the coordinate $x$ in Eq.~(\ref{SinkFFPE}) yields the survival
probability $\mathscr{S}(t)$. Its negative time derivative then delivers the
probability density of first arrival \cite{JPA2003},
\begin{equation}
\wp_{\mathrm{fa}}(t)=-\frac{d}{dt}\int_{-\infty}^{\infty} f(x,t)dx.
\end{equation}

The density function $f(x,t)$ can be determined from Eq.~(\ref{SinkFFPE}) by
application of combined Laplace and Fourier transforms, defined in terms of
\begin{equation}
f(k,s)=\int_0^{\infty}dt\,e^{-st}\int_{-\infty}^{\infty}dx\,e^{ikx}f(x,t).
\end{equation}
The solution reads \cite{JPA2003}
\begin{equation}
f(k,s)=\frac{e^{ikx_0}-\wp_{\mathrm{fa}}(s)e^{ik x_1}}{s+K_{\alpha}|k|^{\alpha}}.
\label{fks1}
\end{equation}
Integrating this result over $k$ yields
\begin{eqnarray}
\int_{-\infty}^{\infty} f(k,s)dk=f(x=0,s)=0,
\end{eqnarray}
and thus
\begin{equation}
W(x_1-x_0,s)-W(0,s)\wp_{\mathrm{fa}}=0,
\end{equation}
where $W(x,t)$ is a solution of Eq.~(\ref{SinkFFPE}) \emph{without\/} the sink
term and reads
\begin{equation}
W(x,s)=\int_{-\infty}^{\infty}\frac{e^{ikx}}{s+K_{\alpha}|k|^{\alpha}}dk
\end{equation}
in Laplace space. Hence the probability of first arrival becomes
\begin{equation}
\wp_{\mathrm{fa}}(s)=\frac{\displaystyle\int_{-\infty}^{\infty}dk\frac{e^{ik(x_1
-x_0)}}{s+K_{\alpha}|k|^{\alpha}}}{\displaystyle\int_{-\infty}^{\infty}dk\frac{1}{
s+K_{\alpha}|k|^{\alpha}}}.
\label{pfa}
\end{equation}
We now use this result together with our definitions Eqs.~(\ref{Pdef}) and
(\ref{Edef}) to assess the random search dynamics by a pure Brownian and LF
searchers for a single target.

\subsection{Brownian search}

If the search is performed by a Brownian searcher in Eq.~(\ref{SinkFFPE}) we
take $\alpha=2$ and the first arrival density can be computed analytically.
In Laplace space it reads
\begin{equation}
\wp_{\mathrm{fa}}(s)=\exp\left(-|x_1-x_0|\sqrt{\frac{s}{K_2}}\right).
\label{pfabrown}
\end{equation}  
Back-transformed, we find in real time that
\begin{equation}
\wp_{\mathrm{fa}}(t)=\frac{|x_1-x_0|}{\sqrt{4\pi K_2t^3}}\exp\left(-\frac{
(x_1-x_0)^2}{4K_2t}\right).
\label{alpha2}
\end{equation}
This is well known L\'evy-Smirnov density \cite{MetzlerOshaninRedner}. The search
reliability (\ref{Pdef}) in this case is $P=1$ and the efficiency reads
\begin{equation}
\mathcal{E}_2=\frac{2K_2}{(x_1-x_0)^2}.
\end{equation}

\subsection{First arrival for L\'evy searcher}

\begin{figure}
\resizebox{0.48\textwidth}{!}{\includegraphics{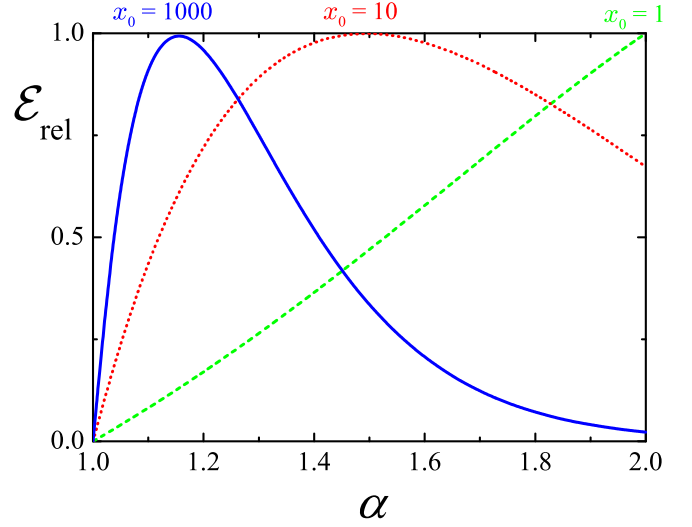}}
\caption{Relative search efficiency for LF search for a single point-like target
as a function of the stable index $\alpha$ according to Eq.~(\ref{EffLevyFlat}),
displayed for the initial searcher-target separations $x_0=1$ (green dashed curve),
$x_0=10$ (red dotted curve), and $x_0=1000$ (blue continuous curve). We take
$K_{\alpha}=1$.}
\label{EffDiffX}
\end{figure}

The first arrival density $\wp_{\mathrm{fa}}(s)$ for an LF searcher in Laplace
space can be computed in terms of Fox' $H$-functions \cite{LevyLong}. From the
small $s$ limit of this function one can see that for $\alpha\le1$ the search
reliability is $P=0$, that is, the search is unsuccessful with probability one,
due to the diverging first absolute moment $\langle|x|\rangle$ of this process.
For $\alpha>1$, $\langle|x|\rangle$ is finite and the reliability is $P=1$. By
integration of the corresponding $H$-function expression in Laplace space one
gets a simple equation for the search efficiency \cite{LevyLong}
\begin{eqnarray}
\mathcal{E}_\alpha=\frac{\alpha K_{\alpha}}{|x_1-x_0|^\alpha}\left|\cos\left(
\frac{\pi\alpha}{2}\right)\right|\Gamma(\alpha),
\label{EffLevyFlat}
\end{eqnarray}
for $1<\alpha<2$. The exact shape of the relative efficiency $\mathcal{E}_{\mathrm{
rel}}=\mathcal{E}_{\alpha}/\mathcal{E}_{\mathrm{opt}}$, where $\mathcal{E}_{\mathrm{
opt}}$ is the maximal (optimal) value of $\mathcal{E}_{\alpha}$ for a given value
of the index $\alpha$, is displayed in Fig.~\ref{EffDiffX}.

\subsection{L\'evy search in the presence of a bias}

\begin{figure}
\resizebox{0.5\textwidth}{!}{\includegraphics{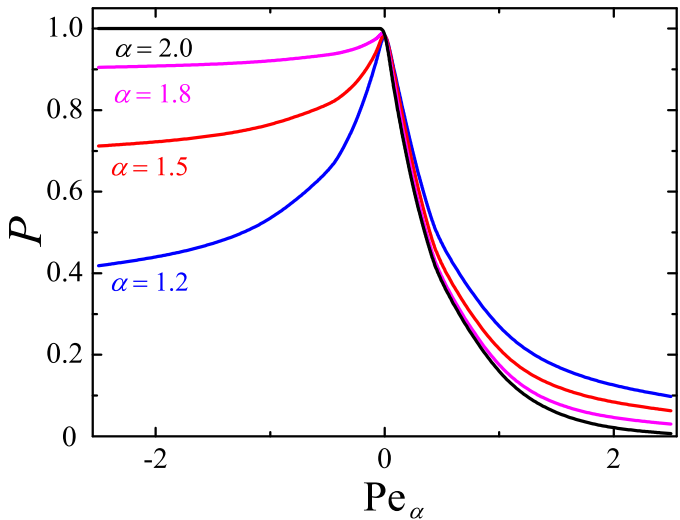}}
\caption{Dependence of the search reliability $P$ for a biased LF on the
generalised P\'eclet number $\mathrm{Pe}_\alpha$, for the indicated values
of the stable index $\alpha$. Positive values of $\mathrm{Pe}_\alpha$
correspond to uphill search.}
\label{Px0}
\end{figure}

The dynamic equation (\ref{SinkFFPE}) can be generalised for search in the
presence of different external potentials. Even a simple bias, stemming from,
for instance, an underwater current for marine predators or wind in the case
of airborne foragers, changes the search performance significantly
\cite{PNAS14,LevyLong}. The space-fractional Fokker-Planck equation then reads
\begin{equation}
\frac{\partial f(x,t)}{\partial t}=K_\alpha\frac{\partial^{\alpha}f(x,t)}{\partial
|x|^{\alpha}}-v\frac{\partial f(x,t)}{\partial x}-\wp_{\mathrm{fa}}(t)\delta(x-x_1).
\label{biasFFPE}
\end{equation} 
Here $v$ denotes the external, constant bias of dimension $[v]=\mathrm{cm}/\mathrm{
sec}$, and the rest of the terms are the same as in Eq.~(\ref{SinkFFPE}). The search
reliability in this case depends solely on a single parameter, the generalised
Pecl\'et number \cite{PNAS14,LevyLong}
\begin{equation}
\mathrm{Pe}_\alpha=\frac{v|x_1-x_0|^{\alpha-1}}{2K_\alpha}.
\end{equation}
In Fig.~\ref{Px0} the search reliability is shown as a function of $\mathrm{Pe}
_\alpha$ for various values of $\alpha$. The target can be either in an uphill
or downhill location relative to the starting point of the searcher. Positive
values of $\mathrm{Pe}_\alpha$ correspond to the uphill scenario, in which the
searcher has to fight the bias in order to reach the target. In this scenario
the reliability increases with decreasing stable index $\alpha$ (as long as
$\alpha>1$). In contrast, for the downhill scenario LFs are less reliable than
Brownian motion, because LFs allow overshoots or leapovers \cite{overshoot} and,
hence, an LF searcher may be eventually lost \cite{LevyLong}. More details about
the search properties by LFs in the presence of an external bias can be found in
Refs.~\cite{PNAS14,LevyLong}.

The exact nature of the external potential landscape creating the bias field
influences the search properties. In Ref.~\cite{Janakiraman17} the fractional
Fokker-Planck equation for search processes was considered for different point
sink strengths for free diffusion, diffusion with a constant bias, and for an
harmonic external potential. A finite strength of the sink describes a finite
probability of absorption. The results for the arrival time density in Ref.
\cite{Janakiraman17} are consistent with our results in \cite{PNAS14,LevyLong}.

\section{Search by a combination of two L\'evy processes}

What happens when we combine two search strategies? This question was analysed
previously in terms of a fractional Fokker-Planck equation for the DNA search
on a long DNA chain in Ref.~\cite{prl2005}, combining Brownian and LF search.
In the language of search processes used here this process was further studied
in Ref.~\cite{JPA16}. In this section we analyse the intermittent motion with
two different LF search strategies governed by the dynamic equation
\begin{equation}
\frac{\partial f(x,t)}{\partial t}=K_{\alpha}\frac{\partial^{\alpha}f(x,t)}{
\partial|x|^{\alpha}}+K_{\mu}\frac{\partial^{\mu}f(x,t)}{\partial|x|^{\mu}}-
\wp_{\mathrm{fa}}(t)\delta(x),
\label{2LevyFFPE}
\end{equation}
which can directly be derived from the corresponding continuous time random walk
model. Here, we introduced the two stable indices $\alpha$ and $\mu$, and the
target position is fixed at $x=0$. The two diffusion coefficients $K_{\alpha}$
and $K_{\mu}$ measure the relative frequency with which jump lengths are drawn
from the corresponding stable laws with indices $\alpha$ and $\mu$. The remaining
variables have the same meaning as before. The first arrival density takes on
the form, similar to (\ref{pfa}),
\begin{eqnarray}
\nonumber
\wp_{\mathrm{fa}}(s)&=&\frac{\displaystyle\int_0^{\infty}\frac{\cos kx_0}{s+
K_\alpha k^{\alpha}+K_\mu k^\mu}dk}{\displaystyle\int_0^{\infty}\frac{1}{s+
K_\alpha k^{\alpha}+K_\mu k^\mu}dk}\\
&=&\frac{\displaystyle\int_0^{\infty}\frac{\cos k}{st_\mu+p k^{\alpha}+k^\mu}
dk}{\displaystyle\int_0^{\infty}\frac{1}{st_\mu+p k^{\alpha}+k^\mu}dk},
\label{pfp2}
\end{eqnarray}
where $t_\mu=|x_0|^\mu/K_\mu$ is a time scale of LFs with the index $\mu$, and
we define
\begin{equation}
p=\frac{K_\alpha}{K_\mu}x_0^{\mu-\alpha}.
\end{equation}

Analogously to the case of a single LF searcher, if both $\alpha>1$ and $\mu>1$
the search reliability is unity, $P=1$ (the motion is recurrent). Vice versa, if
both values are less than or equal to unity, then $P=0$. Thus, only the case
$\alpha<1$ and $\mu>1$ is of interest. The search reliability then reads (see
Appendix \ref{HfuncP}):
\begin{eqnarray}
\nonumber
P&=&\frac{\sin\left(\frac{\pi\left(1-\alpha\right)}{\mu-\alpha}\right)}{2\sqrt\pi}\\
&&\times
H^{12}_{31}\left[\frac{2}{p^\frac{1}{\mu-\alpha}}\begin{array}{|lrrr}\left(1,
\frac{1}{2}\right)\left(\frac{1-\alpha}{\mu-\alpha},\frac{1}{\mu-\alpha}\right)
\left(\frac{1}{2},\frac{1}{2}\right)\\
\left(\frac{1-\alpha}{\mu-\alpha},\frac{1}{\mu-\alpha}\right) \end{array}\right],
\end{eqnarray}
which is a generalisation of Eq.~(16) in Ref.~\cite{JPA16} for the search
reliability of combined L\'evy and Brownian motion.

\begin{figure}
\resizebox{0.5\textwidth}{!}{\includegraphics{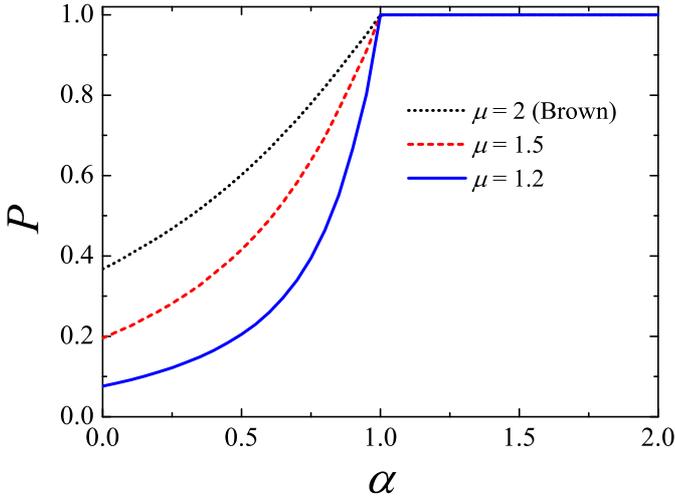}}
\caption{Search reliability $P$ for two combined LF searchers with stable indices
$\alpha$ and $\mu$ as function of $\alpha$, for $p=1$ and different values of
$\mu$..}
\label{alphabetaRel}
\end{figure}

Fig.~\ref{alphabetaRel} illustrates the influence of the competing stable index
$\mu$ on the search reliability $P$, shown as function of $\alpha$. While the
qualitative behaviour for $\mu<2$ stays the same as for Brownian search ($\mu=2$)
\cite{JPA16}, the cumulative probability of finding the target decreases
with $\mu$ once $\alpha<1$. If $\alpha$ is kept constant while $\mu$ is
varied between 1 and 2 then the search reliability increases from 0 to
some value lower than 1 (not shown here). This result can be rephrased in
terms of the classical problem whether a process is recurrent or transient
\cite{hughes}: A recurrent motion revisits the points in the domain of interest,
in our notations it corresponds to $P=1$. We can see that the combination of
recurrent motion ($P=1$ for LFs with $\mu>1$) with transient LFs with $\alpha
\le1$ ($P=0$) leads to a search reliabilities between 0 and 1, that is, the
combined motion is transient with $P>0$. This is one of the central results of
this paper, and it is consistent with our findings for combined L\'evy-Brownian
search \cite{JPA16}, compare also the discussion in Ref.~\cite{prl2005}.

\begin{figure}
\resizebox{0.5\textwidth}{!}{\includegraphics{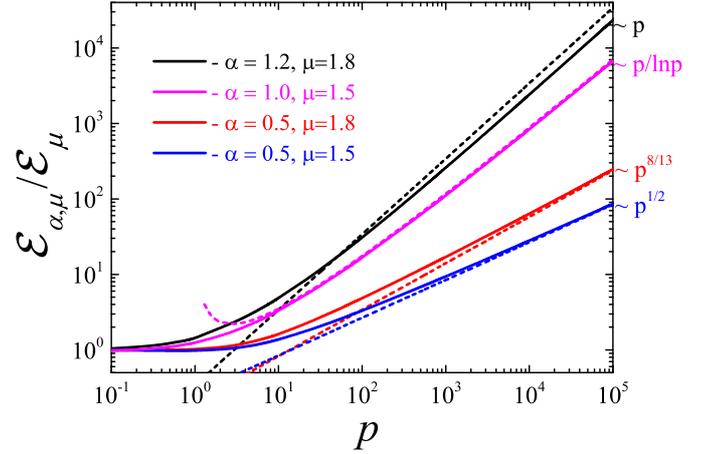}}
\caption{Search efficiency $\mathcal{E}_{\alpha,\mu}$, normalised versus the
single-LF efficiency $\mathcal{E}_{\mu}$ as function of the parameter $p$ for
the search by two LFs. Continuous lines show the numerical results. Dashed
lines represent the corresponding asymptotics derived in Appendix \ref{EffAsympt}.}
\label{EffPmu}
\end{figure}

Following the parametrisation in result (\ref{pfp2}) we plot the search efficiency
$\mathcal{E}_{\alpha,\mu}$ (see Appendix \ref{EffAsympt}) for the dual LF search
as function of the dimensionless parameter $p$ in Fig.~\ref{EffPmu}, for the case
$\alpha<\mu<2$. For $\alpha>1$, $\mathcal{E}_{\alpha,\mu}$ converges to the
efficiency $\mathcal{E}_\alpha$ of a single LF searcher with index $\alpha$ in the
limit $p\to\infty$, that is, $\mathcal{E}_{\alpha>1,\mu}\sim p$. For $\alpha=1$ we
find $\mathcal{E}_{1,\mu}\sim p/\ln p$. These two asymptotics are the same as for
the combined L\'evy-Brownian search \cite{JPA16}. However, for $\alpha<1$ the
asymptotic power law changes to the expression $\mathcal{E}_{\alpha<1,\mu}\sim
p^{(\mu-1)/ (\mu-\alpha)}$. The derivation of these power laws can be found in
Appendix \ref{EffAsympt}.

\begin{figure}
\resizebox{0.47\textwidth}{!}{\includegraphics{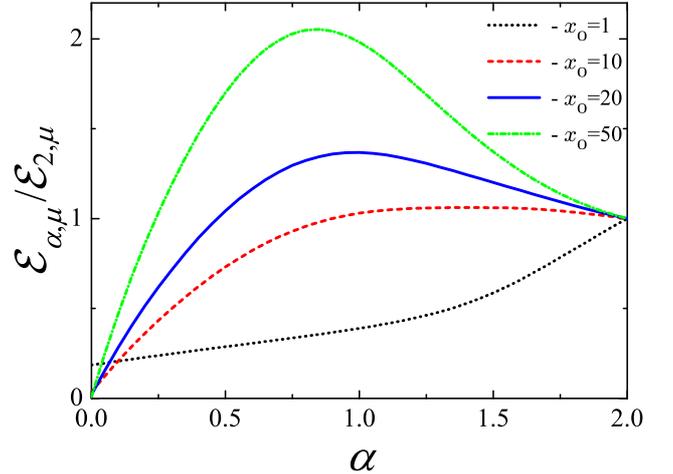}}
\caption{Search efficiency $\mathcal{E}_{\alpha,\mu}$, normalised versus the 
efficiency $\mathcal{E}_{2,\mu}$, of two L\'evy processes with indices $\alpha$ and
$\mu$ as function of $\alpha$ for fixed $x_0$ and $\mu=1.5$. We choose $K_\alpha
=1$ and $K_\mu=1$.}
\label{Effx0fixedmu}
\end{figure}

The comparison of the strategies for different $\alpha$ values and $\mu=1.5$ for
different values of the initial distance $x_0$ is shown in Fig.~\ref{Effx0fixedmu}.
We see that for small $x_0$ the optimal strategy is Brownian ($\alpha=2$), while
for larger $x_0$ it changes to values smaller than 2. This behaviour is analogous
to the combination of Brownian and L\'evy strategies in Ref.~\cite{JPA16}. The
search for a nearby target should be more local in comparison to the search for
far away targets.

\section{Search for two and multiple targets by a single L{\'e}vy searcher}  

In this section we return to the case of a single L{\'e}vy searcher but consider
the situation with multiple, point-like targets. We first consider two targets,
placed at $x_1$ and $x_2$. Then the dynamic equation for the process becomes
(\ref{SinkFFPE})
\begin{eqnarray}
\nonumber
\frac{\partial f(x,t)}{\partial t}&=&K_{\alpha}\frac{\partial^{\alpha}f(x,t)}{
\partial \left\vert x \right\vert^{\alpha}}-\wp_{\mathrm{fa}1}(t)\delta(x-x_{1})-\\
&-&\wp_{\mathrm{fa}2}(t)\delta(x-x_{2}).
\label{2TargFFPE}
\end{eqnarray}
Integrating over the position $x$ it follows, analogously to the above, that
\begin{equation}
\wp_{\mathrm{fa}1}(t)+\wp_{\mathrm{fa}2}(t)=-\frac{d}{dt}\int_{-\infty}^{\infty}
f(x,t)dx.
\end{equation}
The decrease of the survival probability is thus due to the flux into either of
the two targets. Eq.~(\ref{2TargFFPE}) can be solved in Fourier-Laplace space,
producing
\begin{equation}
f(k,s)=\frac{e^{ikx_0}-\wp_{\mathrm{fa}1}(s,x_0)e^{ikx_1}-\wp_{\mathrm{fa}2}(s,
x_0)e^{ikx_2}}{s+K_{\alpha}|k|^{\alpha}}.
\label{fks}
\end{equation}
The inverse Fourier transform can now be taken in the same way as above. As we
here have two conditions of the form $f(x=x_1,s)=f(x=x_2,s)=0$, this inversion
is related to position $x_1$ and $x_2$ of the sinks, and we find the set of
linear equations
\begin{eqnarray}
\nonumber
W(x_1-x_0,s)-\wp_{\mathrm{fa}2}W(x_1-x_2,s)-\wp_{\mathrm{fa}1}W(0,s)=0,\\\nonumber
W(x_2-x_{0},s)-\wp_{\mathrm{fa}1}W(x_2-x_1,s)-\wp_{\mathrm{fa}2}W(0,s)=0.\\
\end{eqnarray}
The density of first arrival is the sum of fluxes to both targets,
\begin{equation}
\wp_{\mathrm{fa}}(s)=\wp_{\mathrm{fa}1}(s)+\wp_{\mathrm{fa}2}(s).
\end{equation}
Let us use the simplified notation
\begin{equation}
W(x_i-x_j,s)=W_{ij}=\int_{-\infty}^{\infty}dk\frac{e^{ik(x_j-x_{i})}}{s+K_{\alpha}
|k|^{\alpha}},
\end{equation}
with $W(0,s)=W_0$. Then the first arrival density becomes
\begin{equation}
\wp_{\mathrm{fa}}(s)=\frac{W_{10}+W_{20}}{W_{12}+W_0}.
\end{equation}
Similarly, the splitting first arrival densities are
\begin{equation}
\wp_{\mathrm{fa}1}(s)=\frac{W_{10}W_0-W_{20}W_{12}}{W_{0}^2-W_{12}^2}
\end{equation} 
and
\begin{equation}
\wp_{\mathrm{fa}2}(s)=\frac{W_{20}W_0-W_{10}W_{12}}{W_{0}^2-W_{12}^2}.
\end{equation}

These expressions can be generalised to the case of multiple targets. The
corresponding dynamic equation reads
\begin{equation}
\frac{\partial f(x,t)}{\partial t}=K_{\alpha}\frac{\partial^{\alpha} f(x,t)}{
\partial|x|^{\alpha}}-\sum_i\wp_{\mathrm{fa}i}(t)\delta(x-x_{i}),
\label{MultipleFFPE}
\end{equation}
where $\wp_{\mathrm{fa}i}(t)$ is the splitting first arrival density to target $i$,
and $x_i$ is the position of target $i$. The formal solution in Fourier-Laplace
space reads
\begin{equation}
f(k,s)=\frac{e^{ikx_0}-\sum_i\wp_{\mathrm{fa}i}(s,x_0)e^{ikx_i}}{s+K_{\alpha}|k|
t^{\alpha}}.
\label{Multiplefks}
\end{equation}
Similarly to the case with two targets this leads to the system of linear equations
\begin{equation}
\wp_{\mathrm{fa}j}W_0+\sum_{i\neq j}W_{ij}\wp_{\mathrm{fa}i}=W_{j0}.
\end{equation}
This system of $n$ equations with $n$ unknowns has a unique solution which allows
one to define all the splitting first arrival densities $\wp_{\mathrm{fa}j}(t)$ as
well as the first arrival density $\wp_{\mathrm{fa}}(t)=\sum_j\wp_{\mathrm{fa}j}
(t)$. The matrix of coefficients in this system of equations is symmetric.
Interestingly, if the targets form an equidistant set, $x_i-x_j=(i-j)\Delta$ with
the constant spacing $\Delta$, the matrix of coefficients is the Toeplitz matrix
\cite{Toeplitz}.

\subsection{Brownian search for two targets}

Let us start with the splitting probabilities for Brownian search. In the
corresponding case $\alpha=2$ \cite{LevyLong}
\begin{eqnarray}
\nonumber
W_{ij}&=&\int_{-\infty}^{\infty}dk\frac{e^{ik(x_i-x_j)}}{s+K_2k^2}\\
&=&\frac{\pi}{\sqrt{sK_2}}\exp\left(-\sqrt{s/K_2}|x_i-x_j|\right).
\end{eqnarray}
Hence,
\begin{equation}
\wp_{\mathrm{fa}}(s)=\frac{e^{-|x_1-x_0|\sqrt{s/K_2}}+e^{ -|x_2-x_0|\sqrt{s/K_2}}}{
e^{-|x_1-x_2|\sqrt{s/K_2}}+1}.
\label{pfagausstwotargets}
\end{equation}
There exist two different cases. In the first case both targets are on the same
side of the starting point ($x_0<x_1<x_2$ or $x_0>x_2>x_1$). In the second case
the starting point is located between the targets ($x_1<x_0<x_2$). 

Let us consider the first case for $x_0<x_1<x_2$. Then,
\begin{eqnarray}
\nonumber
\wp_{\mathrm{fa}}(s)&=&\frac{e^{-(x_1-x_0)\sqrt{s/K_2}}+e^{-(x_2-x_0)\sqrt{s/K_2}}}{
e^{-(x_1-x_2)\sqrt{s/K_2}}+1}\\
&&=e^{-(x_1-x_0)\sqrt{s/K_2}},
\end{eqnarray}
or, after Laplace back transformation,
\begin{equation}
\wp_{\mathrm{pa}}(t)=\frac{x_1-x_0}{2\sqrt{\pi K_2t^3}}\exp\left(-\frac{(x_1-x_0)
^2}{4K_2t}\right).
\end{equation}  
We see that the coordinate $x_2$ of the second target disappears from the expression
of the first arrival probability density and we arrive at the result for a Brownian
particle on a semi-infinite axis, as it should, see Eqs.~(\ref{pfabrown}) and
(\ref{alpha2}): for the Brownian walker first arrival and first passage are
identical, the walker cannot pass the closer target to reach the second target.

In the second case $x_1<x_0<x_2$ from Eq.~(\ref{pfagausstwotargets}) it follows that
\begin{equation}
\wp_{\mathrm{fa}}(s)=\frac{e^{-(x_0-x_1)\sqrt{s/K_2}}+e^{-(x_2-x_0)\sqrt{s/K_2}}}{
e^{-(x_2-x_1)\sqrt{s/K_2}}+1}.
\label{pfagaussbetweentwotargets}
\end{equation}
To compare this solution with the expression for the fluxes in \cite{redner}
(Eqs.~(2.2.10) therein), we note that the latter can be rewritten in our notation
as
\begin{eqnarray}
\nonumber
\wp_{\mathrm{fa}}(s)&=&\wp_{\mathrm{fa}1}(s)+\wp_{\mathrm{fa}2}(s)\\
\nonumber
&=&\frac{\sinh\left((x_2-x_0)\sqrt{\frac{s}{K_2}}\right)+\sinh\left((x_0-x_1)
\sqrt{\frac{s}{K_2}}\right)}{\sinh\left((x_2-x_1)\sqrt{\frac{s}{K_2}}\right)}\\
\nonumber
&=&2\frac{\sinh\left((x_2-x_0)\sqrt{\frac{s}{K_2}}\right)+\sinh\left((x_0-x_1)
\sqrt{\frac{s}{K_2}}
\right)}{(1+e^{-(x_2-x_1)\sqrt{s/K_2}})(e^{(x_2-x_1)\sqrt{s/K_2}}-1)}\\
\nonumber
&=&4\frac{\sinh\left(\frac{(x_2-x_1)\sqrt{\frac{s}{K_2}}}{2}\right)\cosh\left(
\frac{(x_1+x_2-2x_0)\sqrt{\frac{s}{K_2}}}{2}\right)}{(1+e^{-(x_2-x_1)\sqrt{s/
K_2}})(e^{(x_2-x_1)\sqrt{s/K_2}}-1)}\\
&=&\frac{e^{\left(\frac{x_1+x_2}{2}-x_0\right)\sqrt{s/K_2}}+e^{-\left(\frac{x_1+
x_2}{2}-x_0\right)\sqrt{s/K_2}}}{e^{0.5(x_2-x_1)\sqrt{s/K_2}}(1+e^{-(x_2-x_1)\sqrt
{s/K_2}})}.
\end{eqnarray}
The latter expression is equivalent to Eq.~(\ref{pfagaussbetweentwotargets}) after
division of both denominator and numerator by $\exp(0.5(x_2-x_1)\sqrt{s/K_2})$.
For the Brownian case our results thus coincide with those from literature.

\subsection{Long time asymptotics and splitting search reliabilities of an LF
for two targets}

We now derive the splitting probabilities and splitting search reliabilities for
the case LF search for two targets. The values of the search reliabilities can be
found from the asymptotics of $\wp_{\mathrm{fa}}(s)$ in the limit $s\rightarrow0$
(or $t\rightarrow\infty$), for the derivation see Appendix C:
\begin{eqnarray}
\nonumber
\wp_{\mathrm{fa}}(s)&\approx&1-\frac{\Lambda(\alpha)(s/K_{\alpha})^{1-\frac{1}{
\alpha}}}{2}\\
&&\hspace*{-1.4cm}
\times\left(|x_1-x_0|^{\alpha-1}+|x_2-x_0|^{\alpha-1}-|x_2-x_1|^{\alpha-1}\right),
\label{pfaasympttwotarg}
\end{eqnarray}
where
\begin{equation}
\Lambda(\alpha)=\frac{\alpha\Gamma(2-\alpha)}{\pi(\alpha-1)}\sin\left(\frac{\pi
\alpha}{2}\right)\sin\left(\frac{\pi}{\alpha}\right).
\end{equation}
Due to Minkowski's inequality the combination of absolute values in the
brackets of Eq.~(\ref{pfaasympttwotarg}) is always non-negative. From
that expression one can see that the search reliability $P=\wp_{\mathrm{fa}}(
s\rightarrow0)=1$. Now, let us consider the splitting densities. For the first
arrival to the first target (see Appendix C), we find
\begin{eqnarray}
\nonumber
&&\wp_{\mathrm{fa}1}(s)\approx\Big[-|x_1-x_0|^{\alpha-1}+|x_2-x_0|^{\alpha-1}+|x_2
-x_1|^{\alpha-1}\\
&&+\Lambda(\alpha)|x_2-x_0|^{\alpha-1}|x_2-x_1|^{\alpha-1}s^{1-1/\alpha}\Big]/C
\end{eqnarray}
where we use the abbreviation
\begin{equation}
C=2|x_2-x_1|^{\alpha-1}+\Lambda(\alpha)|x_2-x_1|^{2\alpha-2}s^{1-1/\alpha}.
\end{equation}
Then the splitting search reliability to find the first target becomes
\begin{eqnarray}
\nonumber
P_1&=&\frac{-|x_1-x_0|^{\alpha-1}+|x_2-x_0|^{\alpha-1}+|x_2-x_1|^{\alpha-1}}{2|
x_2-x_1|^{\alpha-1}}\\
&&=\frac{1}{2}+\frac{|x_2-x_0|^{\alpha-1}-|x_1-x_0|^{\alpha-1}}{2|x_2-x_1|^{\alpha
-1}}.
\label{P1twotargets}
\end{eqnarray}
Similarly, for the second target
\begin{eqnarray}
\nonumber
P_2&=&\frac{-|x_2-x_0|^{\alpha-1}+|x_1-x_0|^{\alpha-1}+|x_2-x_1|^{\alpha-1}}{2|x_2
-x_1|^{\alpha-1}}\\
&=&\frac{1}{2}-\frac{|x_2-x_0|^{\alpha-1}-|x_1-x_0|^{\alpha-1}}{2|x_2-x_1|^{\alpha
-1}}.
\label{P2twotargets}
\end{eqnarray}
We see that $P_1+P_2=1$ as it should be. Assuming $x_0<x_1<x_2$ we see that for
the Brownian case, $\alpha=2$, the probability $P_2$ equals zero whereas for
$\alpha<2$, $P_2\neq0$. It can also be shown that for $x_0<x_1<x_2$ one always
has $P_2<P_1$. Thus, expressions (\ref{P1twotargets}) to (\ref{P2twotargets})
are a consistent generalisation of the classical result for the splitting
probabilities on an interval (Eq.~(2.2.11) in Ref.~\cite{redner}).

\begin{figure}
\resizebox{0.5\textwidth}{!}{\includegraphics{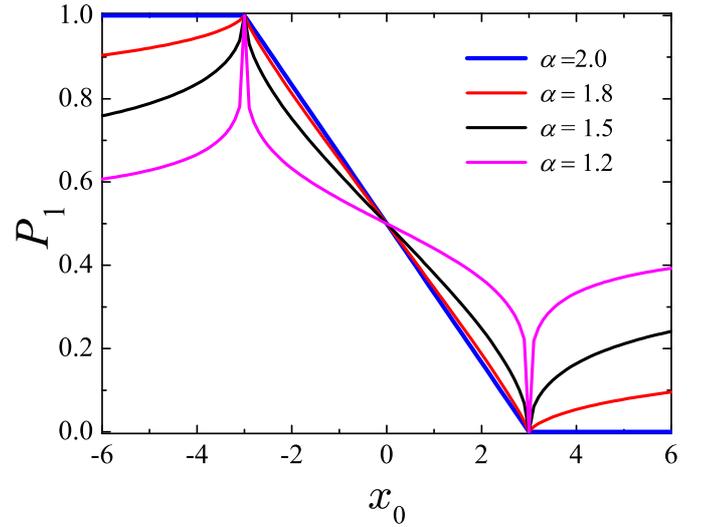}}
\caption{Search reliability $P_1$ of hitting target $x_1$ as a function of
the initial release position $x_0$. The two targets are located at $x_1=-3$
and $x_2=3$, respectively.}
\label{P1x0}
\end{figure}

Fig.~\ref{P1x0} shows the dependence of the search reliability for the first
target $P_1$ as function of the starting position. For the Brownian case the
probability to find the first target is unity if the starting position is to
the left of both targets, zero if the start is to the right, and it changes
linearly with in-between starting points. This behaviour naturally changes for
LF searchers due to the possibility to overshoot the target. The LF searcher
may miss the closest target and hit the one beyond. In the limit of faraway
searcher release, $x_0\to\pm\infty$, the searcher can hit either target with a
likelihood of $\frac{1}{2}$, which can be proven by taking the corresponding
limit in Eqs.~(\ref{P1twotargets}) and (\ref{P2twotargets}). If $\alpha$
decreases from two to unity, that is, when the jump lengths get increasingly
longer tailed while the motion is still recurrent, the probabilities $P_1$
and $P_2$  approach $\frac{1}{2}$. We also see that the search reliability
for hitting the closest target drops with decreasing $\alpha$, while the
chance to hit the further target increases.

\begin{figure}
\resizebox{0.49\textwidth}{!}{\includegraphics{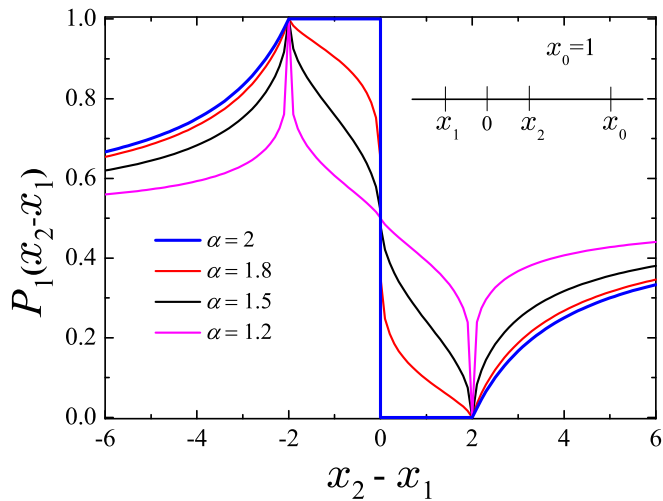}}
\caption{Splitting search reliability $P_1$ for hitting target $x_1$ as function
of $x_2-x_1$. The starting position is fixed at $x_0=1$. The targets at $x_1$
and $x_2$ are placed symmetrically around 0.}
\label{P1x2x1}
\end{figure}

In Fig.~\ref{P1x2x1} we fix the starting position $x_0$ and change the distance
between the targets to show the behaviour of the search reliability $P_1$ (see
the sketch in the inset of the figure). Negative distances correspond to the
inverted order of targets, that is, target 1 is located to the right, while target
2 is the left one. Values of $x_2-x_1=\pm2$ correspond either to the situation when
$x_2$ coincides with $x_0$ or $x_1$ with $x_0$, hence, $P_1(-2)=1$, $P_1(2)=0$
for any $\alpha$. For the Brownian case $\alpha=2$ the splitting search
reliability to find the target $x_1$ is unity when $x_0\leq x_1<x_2$ or $x_0\geq
x_1>x_2$. In the cases $x_0\leq x_2<x_1$ and $x_0\geq x_2>x_1$ the search
reliability to find the target $x_1$ is zero. When the starting position is
located between targets $x_1$ and $x_2$, the splitting search reliability for
the furthermost target increases. In the limit $x_2-x_1\to\infty$ the probability
to hit both targets is the same and tends to $\frac{1}{2}$. Similarly to the
situation in Fig.~\ref{P1x0} we see that LF search always provides the possibility
to hit both targets, and the likelihood for this to happen approaches $\frac{1}{2}$
for $\alpha\rightarrow1$.

\subsection{Search efficiency}

\begin{figure}
\resizebox{0.49\textwidth}{!}{\includegraphics{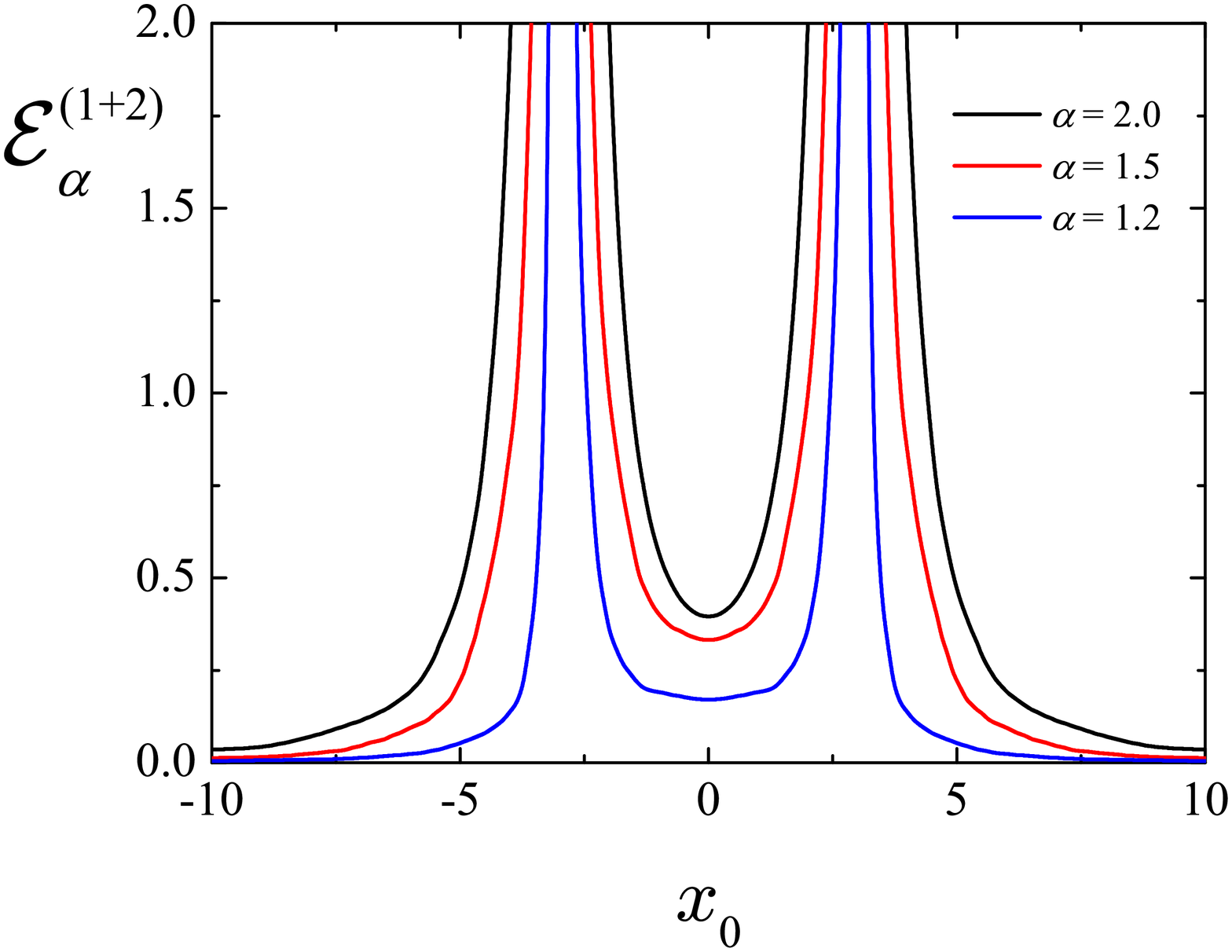}}
\caption{Search efficiency as function of the starting position $x_0$. The targets
are located at $x_1=-3$ and $x_2=3$, respectively.}
\label{Effx0Close}
\end{figure}

\begin{figure}
\resizebox{0.49\textwidth}{!}{\includegraphics{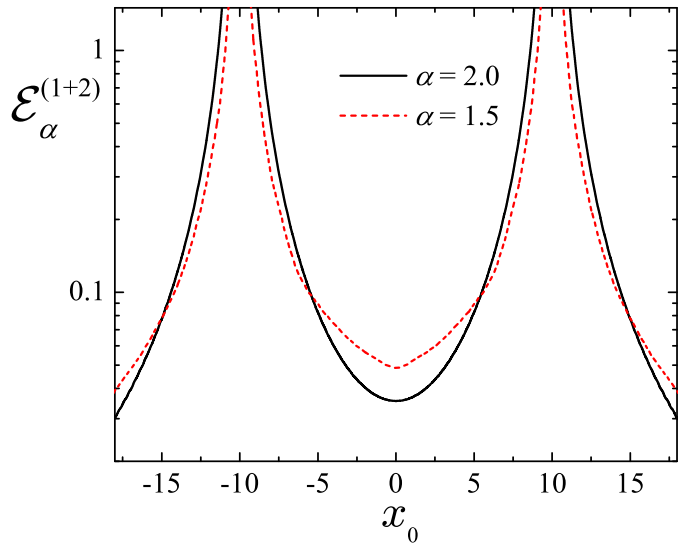}}
\caption{Search efficiency as function of the starting position $x_0$. The targets
are located at $x_1=-10$ and $x_2=10$, respectively.}
\label{Effx0Far}
\end{figure}

\begin{figure}
\resizebox{0.5\textwidth}{!}{\includegraphics{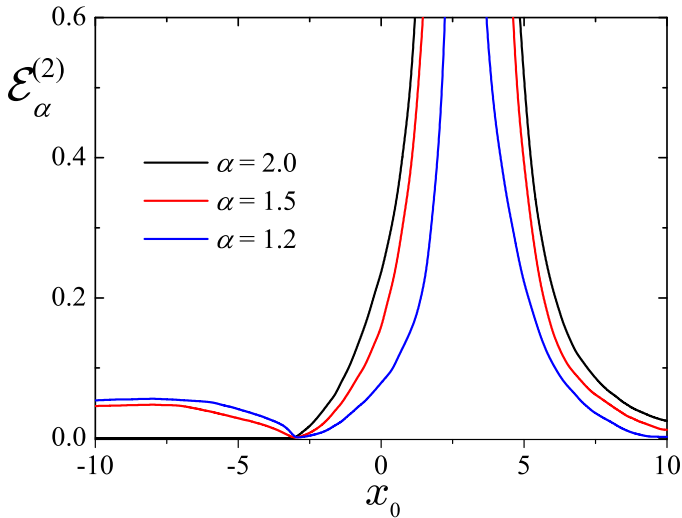}}
\caption{Splitting search efficiency of hitting target $x_2$ as function of
the starting position $x_0$. The targets are located at $x_1=-3$ and $x_2=3$,
respectively.}
\label{Eff2x0}
\end{figure}

In Figs.~\ref{Effx0Close} and \ref{Effx0Far} the target search efficiency is
plotted as a function of the starting position $x_0$. The efficiency exhibits a
symmetry with respect to the midpoint between the targets at positions $x_1$ and
$x_2$. When the starting position is moved towards one of the targets, the search
efficiency rapidly increases. If the starting position is between the targets, the
search efficiency of the Brownian search exceeds that of LFs. However, once the
targets are further apart (Fig.~\ref{Effx0Far}) the midpoint between the targets
is comparatively far from both targets and the LF strategy becomes advantageous
again. In both cases if the searcher starts far to the right or far to the left
from both targets then LF search is more efficient than Brownian. The splitting
search efficiency of hitting a single target, for instance, the second target
(Fig.~\ref{Eff2x0}) behaves similarly to the search efficiency to hit either
target if the searcher starts to the right of the first target. Naturally, if it
starts from the position of the first target, the efficiency of reaching the
second one is 0. We note that if the searcher needs to overshoot one target in
order to reach the second target (the region to the left of the first target in
Fig.~\ref{Eff2x0}), the splitting search efficiency of Brownian motion is 0
because of the absence of overshoots. In contrast LFs are able to do this, and
the splitting search efficiency increases with decreasing $\alpha$.

\begin{figure}
\resizebox{0.49\textwidth}{!}{\includegraphics{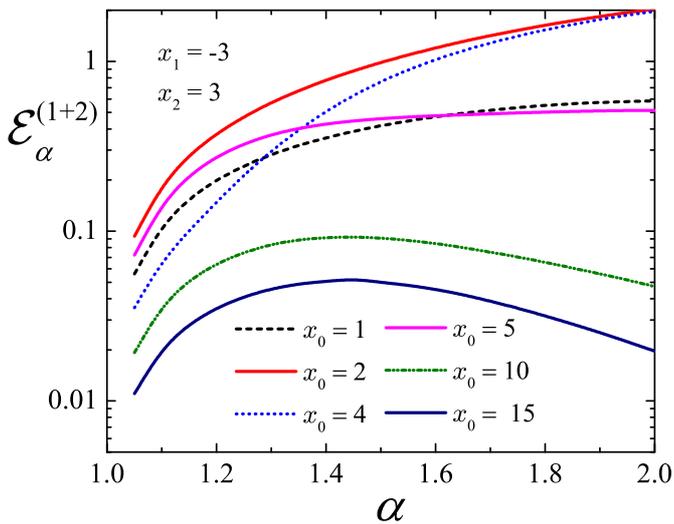}}
\caption{Search efficiency as function of $\alpha$, $x_1=-3$, $x_2=3$. $x_0=1$
and 2 correspond to the starting position between the targets. The rest of the
$x_0$ value are to the right from the right one.}
\label{strategy}
\end{figure}

Search strategies with different $\alpha$ are compared for fixed initial and
target positions in Fig.~\ref{strategy}. As in Ref.~\cite{PNAS14} one can see
that for small distances from at least on of the targets Brownian search is
more efficient than any type of LFs. However, once the searcher starts further
away the longer jumps are a more efficient option. For the same distance $|x_2-
x_0|$ the search efficiency differs, depending on whether one starts on the
same half-line with the other target or not. Surprisingly for some $\alpha$
values the efficiency of target search for $x_1<x_0<x_2$ can be smaller than
for $x_1<x_2<x_0$ for the same $x_2-x_0$ (curves for $x_0=1$ and $x_0=5$ in
Fig.~\ref{strategy}).

\section{Conclusions}

Here we summarised the basic features of LF search for rare targets as well as
developed further generalisations of the LF search model. Thus we considered
search by a combination of two LF search processes and the search for multiple
targets. Some common patterns can be deduced from all these processes. Namely,
in those cases when the target search is easy (for instance, the searcher
starts close to the target or the targets, or the target is in a downhill
position relative to the starting point of the searcher) the pure Brownian
search strategy beats any LF process, or a combined strategy, with respect to
both search reliability and efficiency. However, once the targets are far
away, or the search should go in an upstream direction, LF search with its
substantial probability of long jumps becomes more successful.

Interestingly, the search properties of combined search strategies cannot be
devised from the features of constituent processes. This fact can be clearly
demonstrated if one L\'evy process has a power-law exponent $\alpha\leq1$,
while the second process has an exponent $\mu>1$. The first process alone
is not capable of locating the target independently of how long the search
is, that is, both the search reliability and efficiency are zero. The second
process alone eventually locates the target with certainty (in the absence of
a potential, which could drive a searcher infinitely away from the target),
that is, the second process is recurrent. Non-trivially, the combined process
will have a finite search reliability. Our results also show that this effect
as well as other search properties stay qualitatively the same in the limiting
case when one of the two search processes is Brownian.

As discussed above the generalisation of our approach for the case of many
targets leads to similar conclusions regarding successful search strategies.
However, some new questions appear in this case. One of the most important is
the classical question of splitting probabilities \cite{redner}. In the case
of two targets we generalise a well-known expression for splitting probabilities
of Brownian motion for the case of LF search. The generalised expression reflects
the important difference between the continuous exploration of Brownian motion
and the jump-like behaviour of LFs. The latter leads to leapovers across a target
and, thus, allows the searcher to hit either target from any starting position.

LF search is an idealised process. From a physical point of view the divergence
of the second moment of LFs poses can be avoided by the spatiotemporal coupling
offered by L{\'e}vy walks---at the expense of the relatively straightforward
analytical accessibility. From a biological point of view the assumption that
the jump length statistic is not affected by an external bias may be questionable,
and some penalty in terms of a thermodynamic efficiency concept should be
introduced. However, many of the insights obtained for the simple LF model will,
to some extent, also be present in the L{\'e}vy walk case, for instance, the
question of optimal search for extremely rare targets leads to the same optimal
value for the stable index. In a similar way, we believe that other properties
such as the splitting behaviour will carry over to more complex search processes.
We hope that the results presented here will indeed inspire such research.

\section{Author contributions}

All authors were involved in the preparation of the manu\-script.
All authors have read and approved the final manu\-script.

\begin{acknowledgement}

V.N.M. acknowledges financial support of the RF President Grant for young
scientists (Project no. MD-4550.2016.2). R.M. and A.V.C. acknowledge funding
from the Deutsche Forschungsgemeinschaft through project "Random search
processes, L{\'e}vy flights, and random walks on complex networks".

\end{acknowledgement}

\appendix

\section{Analytical solution for search reliability of two combined LF processes
with stable indices $\alpha\le1$ and $\mu>1$}
\label{HfuncP}

We start with the solution (\ref{pfp2}) of Eq.~(\ref{2LevyFFPE}). If both stable
indices are larger than one, then $P=1$. If both are smaller or equal to one,
then $P=0$. Hence we concentrate here on the case when $\alpha<1$ and $\mu>1$. 
The search reliability for this case reads
\begin{eqnarray}
P=\wp_{\mathrm{fa}}(s=0)=\frac{\int_0^{\infty}\frac{\cos k}{pk^{\alpha}+k^\mu}dk}{
\int_0^{\infty}\frac{1}{p k^{\alpha}+k^\mu}dk}=\frac{I_1}{I_2},
\end{eqnarray}
where
\begin{eqnarray}
I_2=\int_0^\infty\frac{k^{-\alpha}dk}{p+k^{\mu-\alpha}}=\frac{1}{\mu-\alpha}p^{
-\frac{1}{\mu-\alpha}}\frac{\pi}{\sin\left(\pi\frac{1-\alpha}{\mu-\alpha}\right)}
\end{eqnarray}
and
\begin{eqnarray}
\nonumber
I_1&=&\frac{1}{p(\mu-\alpha)}\int_0^\infty k^{-\alpha}\cos k H^{11}_{11}\left[
\frac{k}{p^{\frac{1}{\mu-\alpha}}}\begin{array}{|l}\left(0,\frac{1}{\mu-\alpha}
\right)\\\left(0,\frac{1}{\mu-\alpha}\right)\end{array}\right]dk\\
\nonumber
&=&\frac{\sqrt\pi 2^{-\alpha}}{p(\mu-\alpha)}H^{12}_{31}\left[\frac{2}{p^\frac{
1}{\mu-\alpha}}\begin{array}{|l}\left(\frac{1+\alpha }{2},\frac{1}{2}\right),
\left(0,\frac{1}{\mu-\alpha}\right),\left(\frac{\alpha }{2},\frac{1}{2}\right)\\
\left(0,\frac{1}{\mu-\alpha}\right)\end{array}\right].\\
\end{eqnarray}
Hence:
\begin{eqnarray}
\nonumber
P&=&\frac{\sin\left(\frac{\pi\left(1-\alpha\right)}{\mu-\alpha}\right)}{2\sqrt\pi}\\
&&\times
H^{12}_{31}\left[\frac{2}{p^\frac{1}{\mu-\alpha}}\begin{array}{|l}\left(1,\frac{
1}{2}\right),\left(\frac{1-\alpha}{\mu-\alpha},\frac{1}{\mu-\alpha}\right),\left(
\frac{1}{2},\frac{1}{2}\right)\\\left(\frac{1-\alpha}{\mu-\alpha},\frac{1}{\mu-
\alpha}\right)\end{array}\right].
\end{eqnarray}

\section{Asymptotic behaviour of the search efficiency for $\mu>1$ and $p\to
\infty$.}
\label{EffAsympt}

The search efficiency can be expressed through dimensionless units and the
timescale of the process with stable index $\mu$ as follows
\begin{equation}
\mathcal{E}_{\alpha,\mu}=\frac{1}{t_\mu}\int_0^{\infty}\wp_{\mathrm{fa}}(p,s)
d(st_\mu),
\label{Ep}
\end{equation}
where $t_\mu=x_0^\mu/K_\mu$ is the time scale of LFs with stable index $\mu$.

\subsection{$\alpha>1$}
 
For $\alpha>1$ and $\mu>1$ both LFs have a finite search reliability. In the
limit $p\to\infty$ LFs with exponent $\alpha$ dominate the search process,
hence, the efficiency should take the form of expression (\ref{EffLevyFlat}).

\subsection{$\alpha<1$}

In this case the convergence of expression (\ref{pfp2}) is due to the term $k^\mu$.
Hence the latter cannot be neglected as for $\alpha>1$. Let us change the
variables in Eq.~(\ref{Ep}) as $st_\mu=p^{\nu}u$ and $k=p^\gamma\kappa$, where
$\nu$ and $\gamma$ will be specified below. Then from Eq.~(\ref{Ep}) we get 
\begin{equation}
\mathcal{E}_{\alpha,\mu}t_\mu=p^\nu\int_0^\infty du
\frac{\displaystyle\int_0^{\infty}\frac{\cos (p^{\gamma}\kappa)}{p^\nu
u+p^{1+\alpha\gamma}\kappa^\alpha+p^{\mu\gamma}\kappa^\mu}p^\gamma
d\kappa}{\displaystyle\int_0^{\infty}\frac{1}{p^\nu
u+p^{1+\alpha\gamma}\kappa^\alpha+p^{\mu\gamma}\kappa^\mu}p^\gamma d\kappa}.
\label{split2}
\end{equation}
We choose $\nu$ and $\gamma$ such that
\begin{equation}
\nu=1+\alpha\gamma=\mu\gamma,
\end{equation}
that is,
\begin{equation}
\gamma=\frac{1}{\mu-\alpha}, \qquad\nu=\frac{\mu}{\mu-\alpha}.
\end{equation}
Then Eq.~(\ref{split2}) takes the form
\begin{eqnarray}
\mathcal{E}_{\alpha,\mu}t_\mu=p^\nu\int_0^\infty du \frac{\displaystyle
\int_0^{\infty}\frac{\cos (p^{\gamma}\kappa)}{u+\kappa^\alpha+\kappa^\mu} d\kappa}{
\displaystyle\int_0^{\infty}\frac{1}{ u+\kappa^\alpha+\kappa^\mu}d\kappa}.
\label{split2a}
\end{eqnarray}
The integral in the denominator converges for all positive $u$, does not depend
on $p$, and has an upper bound at $u=0$,
\begin{eqnarray}
\nonumber
f(u)&=&\int_0^{\infty}\frac{1}{ u+\kappa^\alpha+\kappa^\mu}d\kappa\le f(0)\\
&=&\int_0^{\infty}\frac{1}{ \kappa^\alpha+\kappa^\mu}d\kappa.
\label{bound}
\end{eqnarray}
As for the integral in the numerator, since $p\gg1$ the main contribution
comes from small $\kappa$. We thus neglect $\kappa^\mu$ in comparison with
$\kappa^\alpha$ and use the approach from Appendix B in Ref.~\cite{JPA16}.
Hence, for the search efficiency we get
\begin{eqnarray}
\nonumber
\mathcal{E}_{\alpha,\mu}t_\mu&\approx& p^\nu\int_0^\infty \frac{du}{f(u)}
\int_0^{\infty}\frac{\cos (p^{\gamma}\kappa)}{u+\kappa^\alpha} d\kappa\\
&\sim& p^\nu \int_0^\infty \frac{du}{f(u)}\int_0^\infty d\tau e^{-u\tau}\frac{1}{
\tau^{1/\alpha}}l_\alpha\left(\frac{p^\gamma}{\tau^{1/\alpha}}\right),
\end{eqnarray}
where
\begin{eqnarray}
\nonumber
y&=&p^{\gamma}/\tau^{1/\alpha},\\
\nonumber
\tau&=&p^{\gamma\alpha}/y^{\alpha},\\
d\tau=-\frac{1}{\alpha}\frac{p^{\gamma\alpha}}{y^{\alpha+1}}dy.
\end{eqnarray}
Thus 
\onecolumn
\begin{eqnarray}
\nonumber
\mathcal{E}_{\alpha,\mu}t_\mu&\sim& p^{\nu-\gamma+\gamma\alpha}\int_0^\infty
\frac{du}{f(u)}\int_0^\infty dy \exp\left(-\frac{up^{\gamma\alpha}}{y^{\gamma\alpha}}
\right)y^{-\alpha}l_\alpha\left(y\right)\\
\nonumber
&=&p^{\nu-\gamma+\gamma\alpha} \int_0^\infty dyl_\alpha\left(y\right) y^{-\alpha}\int
_0^\infty\frac{du}{f(u)}\exp\left(-\frac{up^{\gamma\alpha}}{y^{\gamma\alpha}}\right)\\
\nonumber
&=&p^{\nu-\gamma} \int_0^\infty dyl_\alpha\left(y\right) \int_0^\infty \frac{e^{-t}
dt}{\displaystyle f\left(\frac{y^\alpha t}{p^{\gamma\alpha}}\right)}\\
&\sim& p^{\nu-\gamma}\sim p^{(\mu-1)/(\mu-\alpha)},\, \mathrm{for}\,\, p\rightarrow\infty,
\end{eqnarray}
due to relation (\ref{bound}), that is, for $\alpha<1$
\begin{equation}
\mathcal{E}_{\alpha,\mu}\sim p^{\frac{\mu-1}{\mu-\alpha}}.
\end{equation}

\subsection{$\alpha=1$}

For $\alpha=1$ we can use Eq.~(\ref{split2a}) with $\nu=\frac{\mu}{\mu-1}$ and
$\gamma=\frac{1}{\mu-1}$, such that
\begin{eqnarray}
\mathcal{E}_{\alpha,\mu}t_\mu=p^\nu\int_0^\infty du \frac{\displaystyle
\int_0^{\infty}\frac{\cos (p^{\gamma}\kappa)}{u+\kappa+\kappa^\mu} d\kappa}{
\displaystyle\int_0^{\infty}\frac{1}{u+\kappa+\kappa^\mu}d\kappa}=p^\nu\int_0
^\infty du\frac{f(u)}{f_1(u)}.
\label{epsalpha1}
\end{eqnarray}
For the integral in the numerator, similar to the case $\alpha<1$ (Appendix B.2)
the main contribution comes from small $\kappa$ values due to $p\gg1$. Hence we can
neglect $\kappa^\mu$ in comparison with $\kappa$. Thus
\begin{eqnarray}
f(u)=\int_0^{\infty}\frac{\cos(p^\gamma\kappa)}{u+\kappa+\kappa^\mu}d\kappa
\simeq\int_0^{\infty}\frac{\cos(p^\gamma\kappa)}{u+\kappa}d\kappa=g(p^\gamma u),
\end{eqnarray}
where $g(z)$ can be expressed through sine and cosine integrals $\mathrm{Si}(z)$ and $\mathrm{Ci}(z)$ as
\begin{eqnarray}
&&g(z)=-\mathrm{Ci}(z)\cos(z)-\left(\mathrm{Si}(z)-\pi/2\right)\sin(z),\\
&&\mathrm{Si}(z)=\int_0^{z}\frac{\sin y}{y}dy,\\
&&\mathrm{Ci}(z)=\int_{z}^\infty\frac{\cos y}{y}dy.
\end{eqnarray}  
The function in the denominator for small arguments $u$ depends on $u$ as $f_1(u)
\sim-\ln u$. Eq.~(\ref{epsalpha1}) can be rewritten in the form
\begin{equation}
\mathcal{E}_{\alpha,\mu}t_\mu=p^\frac{\mu}{\mu-1}\int_0^{1}\frac{du}{f_1(u)}g
\left(p^{\frac{1}{\mu-1}} u\right)+p^\frac{\mu}{\mu-1}\int_{1}^{\infty}\frac{du}{
f_1(u)}g\left(p^{\frac{1}{\mu-1}} u\right).
\label{splitalphaone}
\end{equation}
For the second term in the latter expression one can use the asymptotic of $g(z)
\sim1/z^2$ for $pu\gg1$ since $p\gg1$ (see Eq.~(5.2.35) in Ref.~\cite{Abramowitz}).
This implies that the contribution from the second term decreases with increasing
$p$ at large $p$ values. 

The first term can be rewritten as
\begin{equation}
p^\frac{\mu}{\mu-1}\int_0^{1} \frac{du}{f_1(u)}g(p^{\frac{1}{\mu-1}}u)=p\int_0^{
p^{\frac{1}{\mu-1}}} \frac{dy}{f_1\left(y/p^{\frac{1}{\mu-1}}\right)}g(y).
\end{equation}
The upper bound of this term is given by ($f_1(y)$ is a monotonously decreasing
function of $y$)
\begin{equation}
p\int_0^{p^{\frac{1}{\mu-1}}} \frac{dy}{f_1\left(y/p^{\frac{1}{\mu-1}}\right)}
g(y)<\frac{p}{f_1(1)}\int_{0}^{\infty}dyg(y),
\end{equation}
as $g(y)$ is integrable on $[0,\infty)$ and we can replace the upper limit $p$ of
the integral with $\infty$ at $p\gg1$. Thus, the first term in
Eq.~(\ref{splitalphaone}) does not grow faster than $p$. To get a lower bound for
the growth limit of large $p$ we use the first mean value theorem \cite{Ryzhik}
and a small argument asymptotic $f_1(u)\sim-\ln u$, yielding
\begin{equation}
p\int_0^{p^{\frac{1}{\mu-1}}}\frac{dy}{f_1\left(y/p^{\frac{1}{\mu-1}}\right)}g(y)=\frac{p}{f_1\left(y^{*}/p^{\frac{1}{\mu-1}}\right)}\int_0^{p{\frac{1}{\mu-1}}}
dyg(y)\sim\frac{p}{-\ln\left(y^*/p^{\frac{1}{\mu-1}}\right)}\int_0^\infty dy g(y),
\end{equation}
where $0<y^*<p^{\frac{1}{\mu-1}}$. Hence
\begin{equation}
\mathcal{E}_{\alpha,\mu}t_\mu\sim\frac{p}{\ln p},
\end{equation}
which is confirmed by numerical simulations.

\section{Derivation of long time asymptotics for two targets}

The first arrival density reads
\begin{equation}
\wp_{\mathrm{fa}}(s)=\frac{\frac{W_{10}}{W_0}+\frac{W_{20}}{W_0}}{\frac{W_{12}}{
W_0}+1}.
\end{equation}
For the ratio $W_{ij}/W_0$ the limit of small $s$ was calculated in Appendix A
of Ref.~\cite{JPA16}:
\begin{equation}
\frac{W_{ij}(s)}{W_0(s)}\approx1-\Lambda(\alpha)s^{1-\frac{1}{\alpha}}|x_j
-x_i|^{\alpha-1}
\end{equation}
Correspondingly,
\begin{eqnarray}
\nonumber
&&\wp_{\mathrm{fa}}(s)\approx\frac{2-\Lambda(\alpha)s^{1-\frac{1}{\alpha}}\left(|x_1-x_0|^{\alpha-1}+|x_2-x_0|^{\alpha-1}\right)}{2-\Lambda(\alpha)s^{1-\frac{1}{\alpha}}|x_2-x_1|^{\alpha-1}}\approx\\
\nonumber&&1-\frac{\Lambda(\alpha)}{2}s^{1-\frac{1}{\alpha}}\left(|x_1-x_0|^{\alpha-1}+|x_2-x_0|^{\alpha-1}-|x_2-x_1|^{\alpha-1}\right),
\end{eqnarray}
where
\begin{equation}
\Lambda(\alpha)=\frac{\alpha\Gamma(2-\alpha)}{\pi(\alpha-1)}\sin\left(\frac{\pi
\alpha}{2}\right)\sin\left(\frac{\pi}{\alpha}\right).
\end{equation}
Due to Minkowski's inequality the combination of the absolute values in the
brackets is always non-negative. From that expression one can see that the
search reliability is $P=1$. Now, let us consider the splitting densities.
The probability to hit the first target can be written as
\begin{eqnarray}
\nonumber
&&\wp_{\mathrm{fa}1}(s)=\frac{W_{20}W_{12}-W_{10}W_{0}}{W_{12}^2-W_0^2}=\frac{\left(\frac{W_{20}}{W_0}\right)^2-\frac{W_{10}}{W_{0}}}{\left(\frac{W_{12}}{W_0}\right)^2-1}\approx \\\nonumber&&\approx\frac{\left(1-\Lambda(\alpha)s^{1-\frac{1}{\alpha}}|x_2-x_0|^{\alpha-1}\right)\left(1-\Lambda(\alpha)s^{1-\frac{1}{\alpha}}|x_2-x_1|^{\alpha-1}\right)-\left(1-\Lambda(\alpha)s^{1-\frac{1}{\alpha}}|x_1-x_0|^{\alpha-1}\right)}{\left(1-\Lambda(\alpha)s^{1-\frac{1}{\alpha}}|x_2-x_1|^{\alpha-1}\right)^2-1}\approx\\
&&\approx\frac{-|x_1-x_0|^{\alpha-1}+|x_2-x_0|^{\alpha-1}+|x_2-x_1|^{\alpha-1}+\Lambda(\alpha)|x_2-x_0|^{\alpha-1}|x_2-x_1|^{\alpha-1}s^{1-\frac{1}{\alpha}}}{2|x_2-x_1|^{\alpha-1}+\Lambda(\alpha)|x_2-x_1|^{2\alpha-2}s^{1-\frac{1}{\alpha}}}.
\label{pfa1asymptotics}
\end{eqnarray}
The second splitting probability density $\wp_{\mathrm{fa}2}$ can be computed in
exactly the same way and produces a result, which can be written by swapping
$x_1$ and $x_2$ in the last expression (\ref{pfa1asymptotics}).

\twocolumn


\begin{thebibliography}{110}

\bibitem{mirny2007} G. Kolesov, Z. Wunderlich, O.N. Laikova, M.S. Gelfand and L.A. Mirny, Proc. Natl Acad. Sci. USA, \textbf{104} 13948 (2007).

\bibitem{max2013} M. Bauer and R. Metzler, PLoS One, \textbf{8}, e53956 (2013).

\bibitem{berg} H. C. Berg and E. M. Purcell, Biophys. J. \textbf{20}, 193 (1977).

\bibitem{bialek} W. Bialek and S. Setayeshgar, Proc. Natl. Acad. Sci. USA
\textbf{102}, 10040 (2005).

\bibitem{levine} K. Wang, W. -J. Rappel, R. Kerr, and H. Levine, Phys. Rev. E.
\textbf{75}, 061905 (2007).

\bibitem{aljaz} A. Godec and R. Metzler, Phys. Rev. E \textbf{92}, 010701(R) (2015).

\bibitem{HaBa12} T. Harris, E. Banigan, D. Christian, C. Konradt, E. T.
Wojno, K. Norose, E. Wilson, B. John, W. Weninger,
and A. Luster, Nature 486, 545 (2012).

\bibitem{RFM03} G. Ramos-Fern\'andez, J. L. Mateos, O. Miramontes,
G. Cocho, H. Larralde, and B. Ayala-Orozco, Behav.
Ecol. Sociobiol. 55, 223 (2003).

\bibitem{rescue} M. Becker, F. Blatt, H. Szczerbicka, Multiagent System Technologies, vol. 8076 of Lecture Notes in Computer Science, (Springer Berlin Heidelberg (2013), pp. 19–28),  http://dx.doi.org/10.1007/978-3-642-40776-5\_5

\bibitem{Shles06} M. Shlesinger, Nature 443, 281 (2006).

\bibitem{pavlyukevich} I. Pavlyukevich, J. Comput. Phys., \textbf{226} 1830 (2007). 

\bibitem{Stone07} L. Stone, Theory of Optimal Search (Informs, Hanover,
MD, 2007), 2nd ed.

\bibitem{AG03} S. Alpern and S. Gal, The theory of search games and
rendezvous, International Series in Operations Research
and Managment Science (Kluwer Academic Publishers,
Boston, 2003).

\bibitem{Nath08} R. Nathan, W. M. Getz, E. Revilla, M. Holyoak, R. Kadmon,
D. Saltz, and P. E. Smouse, Proc. Natl. Acad. Sci.,
105, 19052 (2008).

\bibitem{MCB14} V. M\'endez, D. Campos, and F. Bartumeus, Stochastic
Foundations in Movement Ecology, Springer series in synergetics
(Springer, Berlin, 2014).

\bibitem{Vis96} G. Viswanathan, V. Afanasyev, S. Buldyrev, E. Murphy,
P. Prince, and H. Stanley, Nature 381, 413 (1996).

\bibitem{Vis99} G. Viswanathan, S. Buldyrev, S. Havlin, M. da Luz,
E. Raposo, and H. Stanley, Nature 401, 911 (1999).

\bibitem{Edw07} A. Edwards, R. Phillips, N. Watkins, M. Freeman,
E. Murphy, V. Afanasyev, S. Buldyrev, M. da Luz, E. Raposo,
H. Stanley, et al., Nature 449, 1044 (2007).

\bibitem{Sims08} D. Sims, E. Southall, N. Humphries, G. C. Hays, C. J. A.
Bradshaw, J. W. Pitchford, A. James, M. Z. Ahmed,
A. S. Brierley, M. A. Hindell, et al., Nature 451, 1098
(2008).

\bibitem{Sims10} N. Humphries, N. Queiroz, J. Dyer, N. Pade, M. Musy,
K. Schaefer, D. Fuller, J. Brunnschweiler, T. Doyle,
J. Houghton, et al., Nature 465, 1066 (2010).

\bibitem{LICCK12} F. Lenz, T. C. Ings, L. Chittka, A. V. Chechkin, and
R. Klages, Phys. Rev. Lett. 108, 098103/1 (2012).

\bibitem{shlesinger86} M.F. Shlesinger and J. Klafter, 1986, On Growth and Form ed H E Stanley and N Ostrowsky (Dordrecht:
Martinus Nijhoff)

\bibitem{SZK93} M. Shlesinger, G. Zaslavsky, and J. Klafter, Nature 363,
31 (1993).

\bibitem{VLRS11} G. Viswanathan, M. da Luz, E. Raposo, and H. Stanley,
The Physics of Foraging (Cambridge University Press,
Cambridge, 2011).

\bibitem{Pea06} K. Pearson, Biometric ser. 3, 54 (1906).


\bibitem{AmSci1990} W.J. O'Brien, H.I. Browman and B.I. Evans, Am. Sci., \textbf{78}, 152 (1990).

\bibitem{AmerZool2001} D.L. Kramer and R.L. MacLaughlin, Am. Zool., \textbf{41}, 137 (2001).

\bibitem{BenichouIntermittent} O. B\'enichou, C. Loverdo, M. Moreau and R. Voituriez, Rev. Mod. Phys., \textbf{83}, 81 (2011).

\bibitem{benichou2005} O. B\'enichou, M. Coppey, M. Moreau, P.H. Suet and R. Voiturierz, Phys. Rev. Lett., \textbf{94}, 198101 (2005).

\bibitem{benichou2008} C. Loverdo, O. B\'enichou, M. Moreau and R. Voiturierz, Nat. Phys., \textbf{4}, 134 (2008).

\bibitem{ReynoldsPhysA2009} A. Reynolds, Physica A, \textbf{388}, 561 (2009).

\bibitem{gleb1} G. Oshanin, H.S. Wio, K. Lindenberg, and S.F. Burlatsky, J. Phys.: Condens. Matter, \textbf{19}, 065142 (2007).

\bibitem{gleb2} G. Oshanin, K. Lindenberg, H.S. Wio, and S. Burlatsky, J. Phys. A: Math. Theor., \textbf{42}, 434008 (2009).

\bibitem{prl2005} M.A. Lomholt, T. Ambj{\"o}rnsson, and R. Metzler, Phys.
Rev. Lett. \textbf{85}, 260603 (2005).

\bibitem{LomholtPNAS2008} M.A. Lomholt, T. Koren, R. Metzler, and J. Klafter, Proc. Natl. Acad. Sci. U.S.A., \textbf{105}, 11055 (2008).

\bibitem{benichou2006} O. B\'enichou, C. Loverdo, M. Moreau, and R. Voituriez,
Phys. Rev. E, \textbf{74}, 020102 (2006).

\bibitem{benichou2007} O. B\'enichou, C. Loverdo, M. Moreau, and R. Voituriez,
J. Phys. Condens. Matter, \textbf{19}, 065141 (2007).

\bibitem{PNAS14} V. V. Palyulin, A.V. Chechkin and R. Metzler, Proc. Natl. Acad. Sci. USA, \textbf{111}, 2931 (2014).

\bibitem{LevyLong} V. V. Palyulin, A.V. Chechkin and R. Metzler, J. Stat. Mech., P11031 (2014).

\bibitem{redner} S. Redner, 2001, A Guide to First-Passage Processes (Cambridge: Cambridge University Press).

\bibitem{MetzlerOshaninRedner} R. Metzler, G. Oshanin, S. Redner, First-Passage Phenomena and Their Applications, World Scientific, 2014.

\bibitem{JPA2003} A. V. Chechkin, R.Metzler, V. Yu. Gonchar, J. Klafter and
L. V. Tanatarov, J. Phys. A: Math. Gen. \textbf{36}, L537 (2003).

\bibitem{yossinat} S. Condamin, O. B{\'e}nichou, V. Tejedor, R. Voituriez, and J.
Klafter, Nature \textbf{450}, 77 (2007).

\bibitem{chevalier} O. B{\'e}nichou, C. Chevalier, J. Klafter, B. Meyer, and R.
Voituriez, Nat. Chem. \textbf{2}, 472 (2010).

\bibitem{gleb} C. Mejia-Monasterio, G. Oshanin, and G. Schehr, J. Stat. Phys.
\textbf{2011},  P06022.

\bibitem{thiago} T. Mattos, C. Mejia-Monasterio, R. Metzler, and G. Oshanin,
Phys. Rev. E \textbf{86}, 031143 (2012).

\bibitem{aljaz1} A. Godec and R. Metzler, Sci. Rep. \textbf{6}, 20349 (2016).

\bibitem{aljaz2} A. Godec and R. Metzler, Phys. Rev. X \textbf{6}, 041037 (2016).

\bibitem{aldous} Aldous, D. An introduction to covering problems for random walks on graphs. Journal of Theoretical Probability, \textbf{2}, 87-89 (1989).

\bibitem{CBV15} M. Chupeau, O. B\'enichou, and R. Voituriez, Nat. Phys.
11, 844 (2015).

\bibitem{epl} O. B{\'e}nichou, M. Coppey, M. Moreau, and R. Voituriez, Europhys.
Lett. \textbf{75}, 349 (2006).

\bibitem{goat} H.J. de Knegt, G.M. Hengeveld, F. van Langevelde, W.F. de Boer, and K.P. Kirkman, Behav. Ecol., \textbf{18} 1065 (2007).

\bibitem{deer} S. Focardi, P. Montanaro, and E. Pecchioli, PLoS One, \textbf{4}, e6587 (2009).

\bibitem{dinoflagellate} F. Bartumeus, F. Peters, S. Pueyo, C. Marras\'e and J. Catalan, Proc. Natl Acad. Sci., \textbf{100}, 12771 (2003).


\bibitem{brockmann1} D. Brockmann, Phys. World, \textbf{2}, 31 (2010).

\bibitem{brockmann2} D. Brockmann, L. Hufnagel and T. Geisel, Nature, \textbf{439}, 462 (2006).

\bibitem{albatros} A. M. Edwards et al., Nature \textbf{449}, 1044 (2007).

\bibitem{PNAS2012} N. E. Humphries, H. Weimerskirch, N. Queiroza, E. J. Southalla,
and D. W. Sims, Proc. Natl. Acad. Sci. USA \textbf{109}, 7169 (2012).

\bibitem{GnKo54} B. Gnedenko and A. Kolmogorov, Limit distributions for
sums of independent random variables (Addison-Wesley,
1954).

\bibitem{Levy37} P. L\'evy, Theorie De L'addition Des Variables Aleatoires
(Gauthier-Villars, Paris, 1937).

\bibitem{bouchaud} J.-P. Bouchaud and A. Georges, Phys. Rev. \textbf{195}, 127
(2000).

\bibitem{hughes} B.R. Hughes, 1995, Random Walks and Random Environments vol 1 Random Walks (Oxford: Clarendon)

\bibitem{KRS08} R. Klages, G. Radons, and I. Sokolov, eds., Anomalous
transport (Wiley-VCH, Berlin, 2008).

\bibitem{MeKl00} R. Metzler and J. Klafter, Phys. Rep., 339, 1 (2000).

\bibitem{wong} M. F. Shlesinger, J. Klafter, and Y. M. Wong, J. Stat. Phys.
\textbf{27}, 499 (1982).

\bibitem{ZaburdaevReview} V. Zaburdaev, S. Denisov, and J. Klafter, Rev. Mod. Phys., \textbf{87}, 483 (2015). 

\bibitem{KlSo11} J. Klafter and I. Sokolov, First Steps in Random Walks:
From Tools to Applications (Oxford University Press, Oxford, 2011).

\bibitem{Buch08} M. Buchanan, Nature 453, 714 (2008).

\bibitem{dJWH11} M. de Jager, F. J. Weissing, P. M. J. Herman, B. A.
Nolet, and J. van de Koppel, Science, 332, 1551 (2011).

\bibitem{JPE11} A. James, M. J. Plank, and A. M. Edwards, J. Roy. Soc.
Interf. 8, 1233 (2011).

\bibitem{Pyke15} G. Pyke, Meth. Ecol. Evol. 6, 1 (2015).

\bibitem{Reyn15} A. Reynolds, Phys. Life Rev. 14, 59 (2015).

\bibitem{smoluchowski} M. Smoluchowski, Physikal. Zeitschr. \textbf{17}, 557 (1916).

\bibitem{lac} Riggs A.D, Bourgeois S, Cohn M (1970) J Mol Biol 53: 401-417.

\bibitem{adam} G. Adam and M. Delbr{\"u}ck, in Structural Chemistry and Molecular
Biology, edited by A. Rich and N. Davidson (W. H. Freeman, San Francisco, CA, 1968).

\bibitem{richter} P. H. Richter and M. Eigen, Biophys. Chem. \textbf{2}, 255 (1974).

\bibitem{Berg1977} O.G. Berg and C. Blomberg, Biophys. Chem., \textbf{7}, 33 (1977) 

\bibitem{Berg1981} O.G. Berg, R.B. Winter and P.H. von Hippel, Biochemistry, \textbf{20}, 6929 (1981).

\bibitem{Berg1989} P.H. von Hippel and Otto G. Berg, J. Biol. Chem., \textbf{264}, 675 (1989).

\bibitem{BenichouPhysRep2012} M. Sheinman, O. B\'enichou, Y. Kafri and R. Voituriez, Rep. Prog. Phys., \textbf{75}, 026601 (2012).

\bibitem{SlidingExp} H. Kabata, O. Kurosawa, I. Arai, M. Washizu, S.A. Margarson, R.E. Glass, N. Shimamoto, Science, \textbf{262}, 1561-1563
(1993).

\bibitem{WangAustinCoxPRL2006} Y.M. Wang, R.H. Austin, and E.C. Cox, Phys. Rev. Lett., \textbf{97}, 048302 (2006).

\bibitem{mark} I. M. Sokolov, R. Metzler, K. Pant, and M. C. Williams,
Biophys. J. \textbf{89}, 895 (2005).

\bibitem{AssDissExp} Y. Harada, T. Funatsu, K. Murakami, Y. Nonoyama, A. Ishihama, and T. Yanagida, Biophys. J., \textbf{76}, 709-715 (1999).

\bibitem{IntersegExp} C. Bustamante, M. Gutholdi, X. Zhu, and G. Yang, J. Biol. Chem., \textbf{274}, 16665-16668 (1999).

\bibitem{gijs1} B. van den Broek, M. A. Lomholt, S.-M. J. Kalisch, R. Metzler,
and G. J. L. Wuite, Proc. Natl. Acad. Sci. USA \textbf{105}, 15738 (2008).

\bibitem{HalfordMarko2004} S.E. Halford, J.F. Marko, Nucleic Acids Res., \textbf{32}, 3040-3052 (2004).

\bibitem{Benichou2004} M. Coppey, O. B\'enichou, R. Voituriez, and M. Moreau, Biophys J., \textbf{87}, 1640-1649 (2004).

\bibitem{Erskine} S.G. Erskine, G.S. Baldwin, and S.E. Halford, Biochemistry, \textbf{36}, 7567-7576 (1997).

\bibitem{LevyJMB2009} O. Givaty and Y. Levy, J. Mol. Biol., \textbf{385}, 1087-1097 (2009).

\bibitem{Klenin} K.V. Klenin, H. Merlitz, J. Langowski, and C.-X. Wu, Phys. Rev. Lett., \textbf{96}, 018104 (2006).

\bibitem{SlutskyMirny} M. Slutsky and L.A. Mirny, Biophys J., \textbf{87}, 4021-4035 (2004).

\bibitem{gijs2} M. A. Lomholt, B. v. d. Broek, S.-M. J. Kalisch, G. J. L. Wuite,
and R. Metzler, Proc. Natl. Acad. Sci. USA \textbf{106}, 8204 (2009).

\bibitem{shklovskii2007} T. Hu and B.I. Shklovskii, Phys Rev. E, \textbf{76}, 051909 (2007).

\bibitem{LiBerg2009} G.-W. Li, O.G. Berg and J. Elf, Nature Physics, \textbf{5}, 294-297 (2009).

\bibitem{markovitz2013} A. Marcovitz and Y. Levy, Biophys J., \textbf{104}, 2042-2050 (2013).

\bibitem{max2012} M. Bauer, and R. Metzler, Biophys. J., \textbf{102}, 2321-2330 (2012).

\bibitem{tolyaspeedselectivity2013} A. Veksler, and A.B. Kolomeisky, J. Phys. Chem. B, \textbf{117}, 12695 (2013).

\bibitem{larson2013} S. Yu, S. Wang, R. G. Larson, J Biol Phys, \textbf{39},  565-586 (2013).

\bibitem{max_sequence} M. Bauer, E. S. Rasmussen, M. A. Lomholt, and R. Metzler,
Sci. Rep. \textbf{5}, 10072 (2015).

\bibitem{spakowitz} E. F Koslover, M. A. D. Rosa, and A. J. Spakowitz,
Biophys. J. \textbf{101}, 856 (2011).

\bibitem{liu} L. Liu, A. G. Cherstvy, and R. Metzler, J. Phys. Chem. \textbf{121},
1284 (2017).
 
\bibitem{noise} A. Raj and A. van Oudenaarden, Cell \textbf{135}, 216 (2008).

\bibitem{otto} O. Pulkkinen and R. Metzler, Phys. Rev. Lett.  \textbf{110}, 198101
(2013).

\bibitem{HalfordReview2009} S.E. Halford, Biochem. Soc. Trans. \textbf{37}, 343-348 (2009).

\bibitem{finkelstein2012} F. Wang, S. Redding, I.J. Finkelstein, J. Gorman, D.R. Reichman and E.C. Greene, Nat. Struct. Mol. Biol., \textbf{20}, 174-181 (2013). 

\bibitem{berg2012} P. Hammar, P. Leroy, A. Mahmutovic, E.G. Marklund, O.G. Berg, J. Elf, Science, \textbf{336}, 1595-1598 (2012).

\bibitem{mirny2009} L. Mirny, M. Slutsky, Z. Wunderlich, A. Tafvizi, J. Leith and A. Kosmrlj, J. Phys. A: Math. Theor., \textbf{42}, 434013 (2009).

\bibitem{spitzer} F. Spitzer, 1976, Principles of random walk, Springer.

\bibitem{james2010} A. James, J.W. Pitchford, M.J. Plank, Bull. Math. Biol. \textbf{72}, 896 (2010).

\bibitem{PhysRep2012} M. Sheinman, O. B\'enichou, Y. Kafri and R. Voituriez, Rep. Prog. Phys., \textbf{75}, 026601 (2012).

\bibitem{JPA16} V. V. Palyulin, A.V. Chechkin, R. Klages and R. Metzler, J. Phys. A, \textbf{49}, 394002 (2016).

\bibitem{overshoot} T. Koren, M.A. Lomholt, A.V. Chechkin, J. Klafter, and R. Metzler, Phys. Rev. Lett., 99, 160602 (2007).

\bibitem{Janakiraman17} D. Janakiraman, Phys. Rev. E, \textbf{95}, 012154 (2017). 

\bibitem{Toeplitz} E.H. Bareiss, Numerische Mathematik, 13, 404–424 (1969).

\bibitem{Abramowitz} M. Abramowitz and I.A. Stegun, Handbook of Mathematical
Functions with Formulas, Graphs, and Mathematical Tables (NBS, 1964).

\bibitem{Ryzhik} I.S. Gradshteyn, I.M. Ryzhik,
Table of the Integrals, Series and Products (San Diego, CA: Academic Press, 2007).

\end{thebibliography}
\end{document}